\documentclass[a4paper,11pt]{article}

\usepackage{jcappub} %

\usepackage{booktabs} %
\usepackage{siunitx}  %
\usepackage{afterpage, longtable} %
\usepackage{caption} %

\usepackage[normalem]{ulem}
\usepackage{bm}
\usepackage[T1]{fontenc} %
\usepackage{import} %
\usepackage{snapshot} 

\usepackage{textcomp}
\usepackage{bigfoot}

\usepackage[normalem]{ulem}

\usepackage[acronym, toc]{glossaries} 
\glsdisablehyper

\newcommand{\gsim}{\raisebox{-0.13cm}{~\shortstack{$>$ \\[-0.07cm]
      $\sim$}}~}

\title{Millisecond Pulsars in Globular Clusters and Implications for the Galactic Center Gamma-Ray Excess}

\author[a]{Aurelio Amerio,}
\author[b,c]{Dan Hooper,}
\author[d]{and Tim Linden}

\affiliation[a]{Instituto de F\'isica Corpuscular (IFIC), University of Valencia and CSIC, Calle Catedrático José Beltrán 2, 46980 Paterna, Spain}
\affiliation[b]{The University of Wisconsin-Madison, Wisconsin IceCube Particle Astrophysics Center, Madison, WI 53704, USA}
\affiliation[c]{The University of Wisconsin-Madison, Department of Physics, Madison, WI 53704, USA}
\affiliation[d]{The Oskar Klein Centre and Department of Physics, Stockholm University, AlbaNova SE-10691, Stockholm, Sweden}

\abstract{We study the gamma-ray emission from millisecond pulsars within the Milky Way's globular cluster system in order to measure the luminosity function of this source population. We find that these pulsars have a mean luminosity of
$\langle L_{\gamma}\rangle \sim (1-8)\times 10^{33}\, {\rm erg/s}$ (integrated between 0.1 and 100 GeV) and a log-normal width of $\sigma_L \sim 1.4-2.8$. If the Galactic Center Gamma-Ray Excess were produced by pulsars with similar characteristics, Fermi would have already detected $N \sim 17-37$ of these sources, whereas only three such pulsar candidates have been identified. We conclude that the excess gamma-ray emission can originate from pulsars only if they are significantly less bright, on average, than those observed within globular clusters or in the Galactic Plane. This poses a serious challenge for pulsar interpretations of the Galactic Center Gamma-Ray Excess.}

\begin{document}
\maketitle
\flushbottom

\section{Introduction and Background}

A bright and statistically significant excess of GeV-scale gamma-ray emission has been observed from the region surrounding the Galactic Center in the data collected by the Fermi Gamma-Ray Space Telescope~\cite{Goodenough:2009gk,Hooper:2010mq,Hooper:2011ti,Abazajian:2012pn,Hooper:2013rwa,Gordon:2013vta,Daylan:2014rsa,Calore:2014xka,Fermi-LAT:2015sau,Fermi-LAT:2017opo}. This Galactic Center Gamma-Ray Excess (GCE) persists across all existing models of the astrophysical diffuse emission and has not been attributed to any known astrophysical sources or mechanisms~\cite{Cholis:2021rpp,DiMauro:2021raz}. The observed spectrum, angular distribution, and overall intensity of this excess are each consistent with the signal predicted from the annihilation of dark matter particles in the form of a $m\sim 50 \, {\rm GeV}$ thermal relic.

The most prominent astrophysical interpretation of the GCE is that it could be generated by a large population of centrally-located millisecond pulsars (MSPs)~\cite{Hooper:2010mq,Abazajian:2010zy,Hooper:2011ti,Abazajian:2012pn,Gordon:2013vta,Cholis:2014lta,Yuan:2014rca,Petrovic:2014xra,Brandt:2015ula,Malyshev:2024obk,Kuvatova:2024bdn}. This hypothesis is motivated by the fact that pulsars produce gamma-ray emission with a spectral shape that is similar to that of the GCE~\cite{Baltz:2006sv,Hooper:2010mq,Hooper:2011ti,Abazajian:2012pn,Gordon:2013vta}. 

Pulsars are rapidly spinning neutron stars which steadily transfer their rotational kinetic energy into emission at radio, x-ray, and gamma-ray wavelengths. Most pulsars are the product of supernovae, which leave behind a neutron star with a large surface magnetic field ($B\sim 10^{11}-10^{13} \, {\rm G}$) and an initial rotational period on the order of $P \sim 0.1 \, \rm{s}$. Magnetic dipole breaking causes a neutron star to lose its rotational kinetic energy on the following timescale:
\begin{align}
\tau \equiv \frac{E}{\dot{E}} = \frac{P}{2\dot{P}} \approx 1.6 \times 10^5 \, {\rm yr} \times \bigg(\frac{10^{12} \, {\rm G}}{B}\bigg)^2 \bigg(\frac{P}{0.1 \, {\rm s}}\bigg)^2.
\end{align}
Due to their large magnetic fields, young pulsars lose rotational kinetic energy and become faint relatively quickly. In contrast, some neutron stars are ``spun-up'' to high rates of rotation through dynamical interactions with a stellar companion. These ``recycled'' pulsars typically feature periods on the order of $P\sim 2-10$ ms and magnetic fields of $B\sim 10^8-10^9 \, {\rm G}$, allowing them to retain their rotational kinetic energy over much longer timescales, \mbox{$\tau \sim 0.1-100 \, {\rm Gyr}$~\cite{1982Natur.300..728A,Lorimer:2001vd,Phinney:1994gf,Lorimer:2008se,Kiziltan:2009rx}.} The long lifetimes of MSPs make it plausible that large numbers of these objects could be present in the Inner Galaxy, having been produced in situ, or through the dynamical friction and tidal disruption of globular clusters (which are known to contain large populations of MSPs)~\cite{Brandt:2015ula,Fragione:2017rsp,Gnedin:2013cda}.

Pulsar interpretations of the GCE were elevated substantially in 2015, when two independent groups claimed to have identified evidence that the gamma rays associated with the GCE are spatially clustered, as might be expected if this signal is produced by a population of near-threshold point sources~\cite{Lee:2015fea,Bartels:2015aea}. These early analyses, however, have not since held up to scrutiny~\cite{Leane:2019xiy,Leane:2020nmi,Leane:2020pfc,Zhong:2019ycb}. In particular, it has been shown that the non-Poissonian template fit employed in Ref.~\cite{Lee:2015fea} tends to misattribute smooth gamma-ray signals to point source populations, resulting from imperfections in the modeling of the diffuse backgrounds~\cite{Leane:2019xiy,Leane:2020nmi,Leane:2020pfc}, though the effect of these systematic errors on the preference of such models for pulsar interpretations is disputed by Refs.~\cite{Chang:2019ars, Buschmann:2020adf}.  Furthermore, it was demonstrated in Ref.~\cite{Zhong:2019ycb} that when updated point source catalogs are utilized, wavelet-based analysis (such as those employed in Ref.~\cite{Bartels:2015aea}) do
not favor point source interpretations of the GCE. In light of these results, there is now general agreement that the GCE could arise from either a smooth distribution of sources, such as from annihilating dark matter, or from a very large number of faint point sources (for more recent attempts to address this question using machine learning techniques, see Refs.~\cite{List:2021aer,Mishra-Sharma:2021oxe,List:2020mzd,Caron:2017udl, Caron:2022akb, Ramirez:2024oiw}). 

Pulsar interpretations of the GCE were also bolstered by claims that the spatial morphology of the GCE traces known stellar populations, such as the Galactic Bulge and Bar~\cite{Macias:2019omb,Macias:2016nev,Bartels:2017vsx,Ramirez:2024oiw,Song:2024iup}. If confirmed, this would favor an astrophysical origin for this signal. Once again, significant doubt has since been cast on these claims. In particular, several studies have shown that either dark matter-like or bulge-like morphologies can provide a better fit to the gamma-ray data, depending on the background model and masking procedure that is adopted~\cite{Zhong:2024vyi,McDermott:2022zmq,Cholis:2021rpp,DiMauro:2021raz, Song:2024iup}.

Several arguments have been leveled against 
MSP interpretations of the GCE:
\begin{enumerate}
\item{Low-mass X-ray binaries (LMXBs) are the primary progenitors of MSPs~\cite{1982Natur.300..728A,2009Sci...324.1411A}, and yet relatively few of these objects have been observed in the direction of the Inner Galaxy~\cite{Cholis:2014lta}. By comparing the number of LMXBs in the Inner Galaxy to those found in globular clusters, and scaling that ratio to the gamma-ray emission observed from globular clusters, it can estimated that only 4-11\% of the gamma-ray emission associated with the GCE can be attributed to MSPs~\cite{Haggard:2017lyq}.}
\item{Pulsars typically receive natal velocity kicks at the time of their formation, typically on the order $\sim 10^2~{\rm km/s}$. These kicks should cause the spatial distribution of the resulting MSPs to be broader than that of the overall stellar population, and inconsistent with the observed morphology of the GCE, especially regarding the intensity of the excess observed within the innermost $\sim 1^{\circ}$ around the Galactic Center~\cite{Boodram:2022lpn}.}
\item{Very few pulsars have been detected in the vicinity of the Inner Galaxy. This allows us to place strong constraints on the gamma-ray luminosity function of any MSP population that could be responsible for the GCE. These constraints indicate that if MSPs generate the gamma-ray excess, those pulsars must be systematically less bright than those found in other environments. This, in turn, requires a larger number of MSPs to generate the observed intensity of the GCE, on the order of $N_{\rm GCE, tot} \gsim 10^5$~\cite{Holst:2024fvb} (see also, Refs.~\cite{Cholis:2014noa,Hooper:2016rap,Hooper:2015jlu,Bartels:2018xom,Ploeg:2017vai,Ploeg:2020jeh,Dinsmore:2021nip}).}
\end{enumerate}

It has been suggested that these challenges could be mitigated if a large fraction of the Inner Galaxy's MSP population is not formed through the transfer of angular momentum to neutron stars, but instead through the accretion-induced collapse of O-Ne white dwarfs~\cite{Gautam:2021wqn}. Unlike ordinary MSPs, this hypothetical pulsar population would not be evolutionary connected to LMXBs, and their members would not be expected to receive large natal kicks, allowing them to evade the first two of the arguments listed above. It is not at all clear, however, why a large fraction of MSPs in the Inner Galaxy would be produced through accretion induced collapse when this is not the case among pulsar populations observed in other environments. 
In fact, it has generally
been found that accretion-induced collapse only rarely produces binary configurations that are amenable to further mass-transfer-based spin-up of the compact object \cite{Kalogera_1996, Willems_2004} and moreover produces neutron stars with magnetic fields too
small to produce MSP activity \cite{Wijers_1997, sutantyo2000formationbinarymillisecondpulsars}.

Alternatively, if the magnetic fields of MSPs formed through accretion-induced collapse are of the same magnitude as MSPs formed through standard means, there would be no reason to expect either of these two populations to be systematically fainter than the other.

The population of MSPs observed in the Galactic Plane contains many pulsars that would have been detected by Fermi if they had been present in the Inner Galaxy. In particular, if the GCE is produced by MSPs with the same luminosity function and other characteristics as those observed within the Galactic Plane, $N_{\rm MSP} \sim 10-35$ of the pulsars associated with the Inner Galaxy population should have been been detected by Fermi and included within their 4FGL and 3PC source catalogs~\cite{Holst:2024fvb,Fermi-LAT:2023zzt}. As only three such candidate pulsars exist~\cite{Holst:2024fvb}, such a scenario is in significant tension with this possibility.

It remains possible, however, that the Inner Galaxy could contain a large number of MSPs that are systematically fainter or otherwise more difficult to identify than those observed within the Galactic Plane. Motivating this possibility is the fact that the Galactic Bulge experienced a high rate of star formation when it was young, allowing for the formation of a large number of MSPs which are now very old, potentially causing them be systematically lower in luminosity than those found in the Galactic Plane. The stellar populations present within globular clusters are also very old, however, comparable in age to that of the Galactic Bulge. It is important, therefore, to measure the gamma-ray luminosity function of MSPs in globular clusters, as this could have great bearing on the viability of pulsar interpretations of the GCE. 

In this study, we revisit our previous work~\cite{Hooper:2016rap} attempting to constrain the gamma-ray luminosity function of MSPs in globular clusters. To this end, we use the publicly available Fermi data to measure or constrain the gamma-ray emission from each of the Milky Way's 157 globular clusters (for related work, see Refs.~\cite{Hou:2024wpr, ballet2024fermilargeareatelescope,Zhang:2023gwr,Evans:2022zno,Zhang:2022jlx,Yuan:2022egs,deMenezes:2018ilq,Cholis:2014noa,Fermi-LAT:2010gpc}). We detect statistically significant gamma-ray emission from 56 of these globular clusters, 8 of which are not contained in previous
gamma-ray source catalogs. We then use this data, in combination with the visible luminosities and previously calculated stellar encounter rates of the globular clusters, to constrain the gamma-ray luminosity function of the MSPs in these systems. We find that this MSP population has a luminosity function that is consistent with that previously reported for MSPs in the Galactic Plane, featuring a mean luminosity of $\langle L_{\gamma}\rangle \sim (1-8) \times 10^{33} \, {\rm erg/s}$. If the GCE is generated by pulsars with the same gamma-ray luminosity function and other characteristics as those found in globular clusters, Fermi's source catalogs would have contained $N_{\rm MSP} \sim 17-37$ pulsars associated with this population. From this, we conclude that the GCE could be generated by pulsars only if they are significantly lower in luminosity or otherwise more difficult to identify than those in globular clusters (or in the Galactic Plane).

\section{Fermi Data Analysis}
\label{sec:fermi-data-analysis}

In this section, we describe our procedure for measuring the spectrum and intensity of the gamma-ray emission from each of the Milky Way's 157 globular clusters~\cite{harris2010newcatalogglobularclusters}.
We employ the python package, Fermipy~\cite{wood2017fermipyopensourcepythonpackage}, as provided by the Fermi collaboration, to process the data collected by the Fermi-LAT. 
We work with HEALPix maps of order 11, corresponding to approximately $0.03^{\circ}$ bins, and utilize 15.8 years of data.\footnote{MET range:  239557414 - 733276805} We consider photons in the range of 100 MeV to 100 GeV, divided into 5 logarithmic bins per decade. We adopt the standard event selection for gamma-ray point sources, using the \verb|SOURCE| event class and \verb|FRONT| + \verb|BACK| event types. We also employ a zenith angle cut of $90^{\circ}$ and a rocking angle cut of $52^{\circ}$ in order to reduce the contribution from the Earth's limb. We require data quality=1 and only consider data that was taken while in survey mode. We do not use data that was collected while passing through the South Atlantic Anomaly. These and other aspects of our analysis are summarized in Table~\ref{tab:settings}.

\begin{table}[t]
\centering
\begin{tabular}{ | l| l|  }
\hline
HEALPix map order& 11  \\ 
Weeks & $9-826$  \\ 
Emin & 100 MeV \\
Emax & 100 GeV \\
Instrument Response Functions (IRFs) & \verb|P8R3_SOURCE_V3| \\
EVCLASS & 128 (Source)  \\ 
EVTYPE & 3 (Front + Back)  \\
ZMAX & $90^{\circ}$ \\
Rocking angle cut & $52^{\circ}$ \\
Galactic diffuse emission template & \verb|gll_iem_v07| \\
Isotropic background template & \verb|iso_P8R3_SOURCE_V3_v1| \\
Source catalog & \verb|gll_psc_v35| \\ 
\hline
\end{tabular}
\caption{\texttt{Fermi Science Tools} settings used in this analysis. }
\label{tab:settings}
\end{table}

For each globular cluster, we study a $14^{\circ} \times 14^{\circ}$ region-of-interest (ROI) centered around the nominal position of the source. We then fit the energy spectrum of the globular cluster, adopting a parameterization in the form of a power law with an exponential cut-off:
\begin{equation}
    \phi(E) = N_0 \, E^{-\alpha} \, e^{-E/E_c}. 
\label{eq:energy-spectrum}
\end{equation}
In fitting these sources, we follow the standard procedure used in constructing Fermi-LAT source catalogs. 
We include in our fit any sources in the 4FGL-DR4 catalog that are located within in ROI~\cite{ballet2024fermilargeareatelescope}, excluding any globular cluster, pulsar, or unassociated gamma-ray source that is located within $0.2^\circ$ of the fiducial position of the globular cluster in question. We then add a new point source in the position of the globular cluster, as specified in Ref.~\cite{harris2010newcatalogglobularclusters}, and reoptimize the model of the known sources within the ROI.
Following standard procedure, we run the source localization pipeline, readjust the centroid of the globular cluster, and look for new point sources with a power-law spectrum. 
We then move to the characterization of the globular cluster, performing a 3D likelihood scan in the ($\alpha$, $E_{c}$, $N_0$) parameter space. 
During this procedure, we model the Galactic diffuse emission and isotropic background contributions according to the templates provided by the Fermi collaboration (see Table \ref{tab:settings}). We allow their normalization to float, as well as the energy spectrum normalization of any other resolved source within 3 degrees.
This allows us to identify, for each value of the integrated flux, the combination of these parameters that yields the greatest likelihood. 

In Table~\ref{tab:high_TS}, which we report in Appendix \ref{sec:app-tables}, we list each of the 56 globular clusters for which our analysis identified a statistically significant test statistic, \mbox{TS > 25}, along with the best-fit values for $\alpha$, $E_c$, and the total flux (integrated between 0.1 and 100 GeV). Note that our previous study, which utilized 7 years of Fermi data, detected only 25 globular clusters at a level of \mbox{TS > 25}~\cite{Hooper:2016rap}. We have compared our results to those reported in the 4FGL catalog~\cite{ballet2024fermilargeareatelescope} and find them to be in overall good agreement. The only significant differences are found in the cases of NGC 5139, Palomar 6, Terzan 5, and NGC 6218, for which our analysis yields integrated fluxes which are larger than those given in the 4FGL catalog by $2-3\sigma$. 

\noindent
This appears to be due to the spectral parameterization we have adopted, as these four sources are fit in the 4FGL to a log-parabola or power-law parameterization, rather than that of a power-law with an exponential cutoff.

In addition to the 56 sources with TS > 25 listed in Table~\ref{tab:high_TS}, we found an additional 7 sources which yielded TS > 25, but that had their highest TS values in bins below 0.4 GeV. Due to the large point spread function and related uncertainties that are encountered at these low energies, we do not consider these sources (which are listed in Table~\ref{tab:low_peak_TS}) to be unambiguous detections, and only report upper limits on their flux.

Finally, in Table~\ref{tab:low_TS}, we report upper limits on the gamma-ray fluxes from each of the remaining globular clusters considered in our analysis, each of which yielded TS < 25. 

\section{Millisecond Pulsars in Globular Clusters}
\label{sec:cases}

Due to the high stellar densities found within globular clusters, dynamical interactions between stars take place in these systems much more frequently than they do among stars in the Galactic Plane. As a result, large numbers of low-mass X-ray binaries (LMXBs) and millisecond pulsars (MSPs) are each found within globular clusters. At present, a total of 340 MSPs have been identified within 45 different Milky Way globular clusters (see \url{https://www3.mpifr-bonn.mpg.de/staff/pfreire/GCpsr.html}, and Refs.~\cite{Manchester:2004bp,Fermi-LAT:2023zzt}). The majority of these pulsars have only been observed at radio wavelengths, as they are too distant to be detected by Fermi. In fact, Fermi's Third Pulsar Catalog~\cite{Fermi-LAT:2023zzt} contains only four MSPs that are located within a globular cluster: 
PSR J1717+4308A in NGC 6341, PSR J1823-3021A in NGC 6624, J1824-2452A in NGC 6626, and PSR J1835-3259B in NGC 6652. 

The number of MSPs within a given globular cluster is correlated with that cluster's overall visible-band luminosity, $L_V$ (a proxy for stellar mass), and stellar encounter rate, $\Gamma_e$. 

\noindent
The stellar encounter rate within a globular cluster is given by~\cite{2003ASPC..296..245V,1987IAUS..125..187V}
\begin{align}
\Gamma_e = \frac{4\pi}{\sigma_c} \int \rho^2(r) r^2 dr,
\end{align}
where $\rho(r)$ is the stellar density profile of the cluster and $\sigma_c$ is the velocity dispersion near its core. Throughout this analysis, we make use of the stellar encounter rates tabulated in Ref.~\cite{Bahramian:2013ihw} (and as we list in Table~\ref{tab:tab7-9-hopper-linden}).

In this study, we will consider the following five hypotheses for estimating the number of MSPs that will, on average, be found within a given globular cluster:
\begin{enumerate}
    \item {A case in which the mean number of MSPs in a globular cluster is proportional to its stellar encounter rate, $\langle N_{\rm MSP} \rangle = R_{\rm MSP} \, \Gamma_e$. This approximate scaling might be expected if MSPs in globular clusters are formed largely through stellar interactions.}
    \item{A case in which the relationship between the stellar encounter rate and the expected number of pulsars within a globular cluster is not linear, but instead scales as $\langle N_{\rm MSP} \rangle = R_{\rm MSP} \, \Gamma_e^{0.7}$. Empirically speaking, this hypothesis is preferred over that of a strictly linearly proportional relationship~\cite{deMenezes:2018ilq,Feng:2023htd}.}
    \item{A case in which the mean number of MSPs in a globular cluster is proportional to its visible luminosity, $\langle N_{\rm MSP} \rangle = R_{\rm MSP}\, L_V$.}
    \item{A case in which the mean number of MSPs in a globular cluster is related to a combination of its stellar encounter rate and visible luminosity, $\langle N_{\rm MSP} \rangle = R_{\rm MSP} \, \Gamma_e+ R'_{\rm MSP} \, L_V$.}
    \item{A case in which the mean number of MSPs in a globular cluster is related to a combination of a non-linear scaling to its stellar encounter rate and its visible luminosity, $\langle N_{\rm MSP} \rangle = R_{\rm MSP} \, \Gamma_e^{0.7}+ R'_{\rm MSP} \, L_V$.}
\end{enumerate}
For further discussion regarding the dynamics of globular clusters and empirical relationships between their observed parameters, see Refs.~\cite{Feng:2023htd,Ye:2022yxt,deMenezes:2018ilq,Fragione:2018jxd}.

\begin{figure}[t]
\centering
\includegraphics[width=0.49\textwidth]{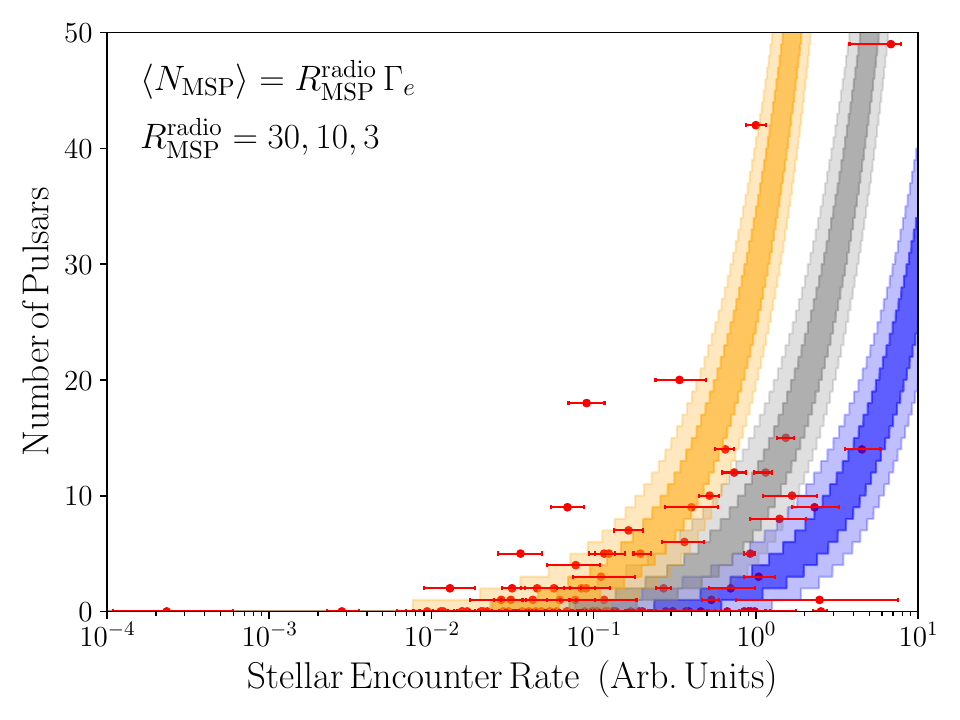}
\includegraphics[width=0.49\textwidth]{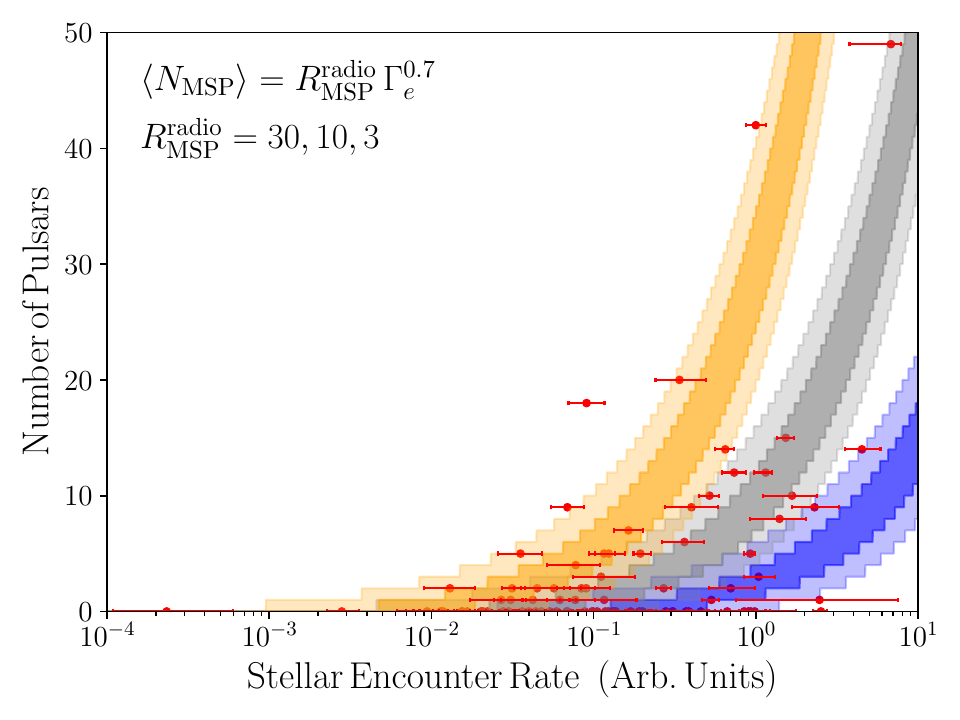}
\includegraphics[width=0.49\textwidth]{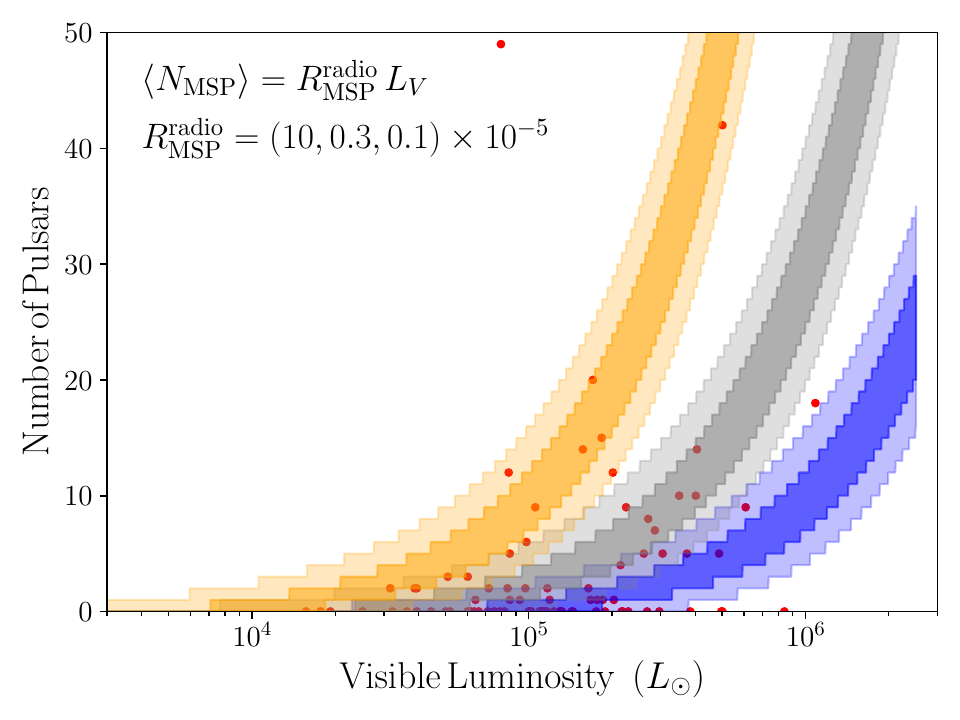}
\includegraphics[width=0.49\textwidth]{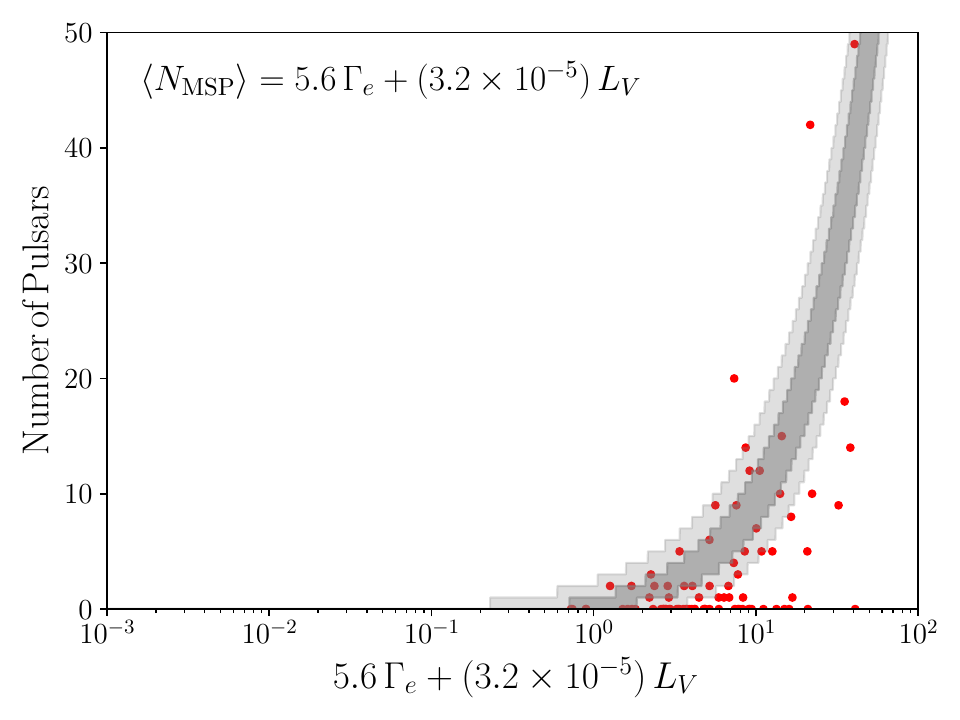}
\includegraphics[width=0.49\textwidth]{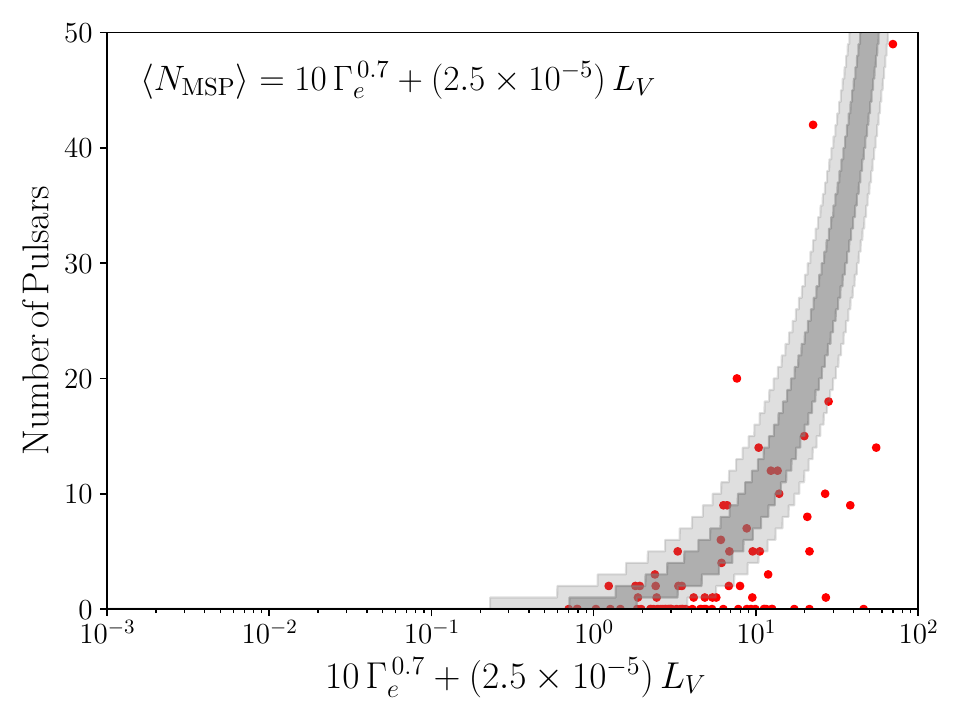}
\caption{The number of known millisecond pulsars in each of the 87 globular clusters considered in this study (those with $L_V > 10^5 \, L_{\odot}$ or $\Gamma_e > 0.01$) as a function of combinations of the stellar encounter rate, $\Gamma_e$, and visible luminosity, $L_V$ (in units of solar luminosities). Also shown are colored bands representing the predicted number of MSPs (68.27\% and 95.45\% containment) in each case, for selected values of $R_{\rm MSP}^{\rm Radio}$. The stellar encounter rate is normalized such that $\Gamma_e =1$ for NGC 104.}
\label{fig:radio}
\end{figure}

\enlargethispage{1\baselineskip} %
Throughout the remainder of this study, we will focus on the 87 globular clusters that have either $L_V > 10^5 \, L_{\odot}$ or $\Gamma_e > 0.01$ (in units such that $\Gamma_e =1$ for the case of NGC 104). In Fig.~\ref{fig:radio}, we plot the number of MSPs that are known to exist in each of these 87 globular clusters, as a function of their stellar encounter rate, visible luminosity, or combinations of these parameters. We also show bands which represent the predicted relationships between these quantities, for selected values of $R_{\rm MSP}^{\rm Radio}$.\footnote{While we take $R_{\rm MSP}$ and $R'_{\rm MSP}$ to be related to the predicted number of gamma-ray MSPs in a given globular cluster, the same quantities with the superscript ``Radio'' correspond to the predicted number of radio MSPs. Due to the different beaming angles of pulsars in the gamma-ray and radio bands, these quantities are not necessarily the same.} 
The inner (outer) bands are calculated such that they cover 68.27\% (95.45\%) of a Poisson distribution for a mean value of $\langle N_{\rm MSP}^{\rm Radio} \rangle$. Even by eye, it is clear that the correlation is less tight for the case of $\langle N_{\rm MSP} \rangle = R_{\rm MSP}\,{\Gamma_e}$ than for $\langle N_{\rm MSP} \rangle = R_{\rm MSP}\,{\Gamma_e}^{0.7}$, or for $\langle N_{\rm MSP} \rangle = L_V$. The tightest correlations are found for combinations of $\Gamma_e$ and $L_V$ (as shown in the final two frames). One should bear in mind that the current catalog of the MSPs in globular clusters is incomplete, limiting the utility of any quantitative comparisons. That being said, the results presented in Fig.~\ref{fig:radio} provide us with motivation to consider the five hypotheses described above for the anticipated number of MSPs in globular clusters.

\section{The Millisecond Pulsar Gamma-Ray Luminosity Function}
\label{sec:glf}

For the gamma-ray luminosity function of MSPs, we adopt a log-normal parameterization:
\begin{align}
\frac{dP}{dL_{\gamma}}(L_{\gamma}) = \frac{1}{L_{\gamma} \sigma_L \sqrt{2\pi}} \exp\bigg(-\frac{(\ln L_{\gamma} - \ln L_0)^2}{2 \sigma_L^2}\bigg),
\label{eq:lumfunc}
\end{align}
where $L_0$ and $\sigma_L$ characterize the center and width of this distribution, respectively. We will often refer to this luminosity function in terms of the mean luminosity, which is given by $\langle L_{\gamma} \rangle = L_0 \, e^{\sigma_L^2/2}$.

The luminosity function of a globular cluster that contains a single MSP is simply equal to the expression given in Eq.~\ref{eq:lumfunc},
\begin{align}
\frac{dP}{dL_{\rm tot}} (L_{\rm tot}, N_{\rm MSP}=1) = \frac{dP}{dL_{1}}(L_1=L_{\rm tot}). 
\end{align}
For a cluster with two MSPs, $N_{\rm MSP}=2$, it is instead given by
\begin{align}
\frac{dP}{dL_{\rm tot}}(L_{\rm tot}, N_{\rm MSP}=2) =  \int_0^{L_{\rm tot}} \frac{dP}{dL_1}(L_1) \, \frac{dP}{dL_2}(L_2=L_{\rm tot}-L_1) \, dL_1,
\end{align}
and for $N_{\rm MSP}=3$,
\begin{align}
\frac{dP}{dL_{\rm tot}}(L_{\rm tot}, N_{\rm MSP}=3)  = \int_0^{L_{\rm tot}-L_1} \int_0^{L_{\rm tot}} \frac{dP}{dL_1}(L_1) \, \frac{dP}{dL_2}(L_2) \,\frac{dP}{dL_3}(L_3=L_{\rm tot}-L_1-L_2) \, dL_1 \, dL_2,
\end{align}
and so on.

For a globular cluster containing a given mean number of expected MSPs, $\langle N_{\rm MSP}\rangle$, its gamma-ray luminosity function will be given by
\begin{align}
\frac{dP}{dL_{\rm GC}}(L_{\rm GC},\langle N_{\rm MSP}\rangle)  = \sum_{N_{\rm MSP}=0}^\infty P(N_{\rm MSP}) \, \frac{dP}{dL_{\rm tot}}(L_{\rm tot}, N_{\rm MSP}),
\label{eq:glf}
\end{align}
where 
\begin{align}
P(N_{\rm MSP},\langle N_{\rm MSP}\rangle)= \frac{\langle N_{\rm MSP}\rangle^{N_{\rm MSP}} \, e^{-\langle N_{\rm MSP}\rangle}}{N_{\rm MSP}!}
\end{align}
is the Poisson probability that the globular cluster will contain $N_{\rm MSP}$ pulsars.

In order to constrain the gamma-ray luminosity function of MSPs in globular clusters, we closely follow the analysis from Ref.~\cite{Hooper:2016rap}. 
We repeat these steps here for clarity:
\begin{enumerate}
    \item For a single globular cluster, $i$, we fit the energy spectrum to the Fermi-LAT data, as detailed in Section \ref{sec:fermi-data-analysis}. This gives us the observational likelihood profile for the flux, $\phi_i$, which we can write as $\mathcal{L}_{obs}(\phi_i)$. This function quantifies the measurement uncertainty of the flux.
    \item  We define the theoretical model for the gamma-ray luminosity function as in Eq.~\ref{eq:glf}. This is a probability density function  for the luminosity, $L_{{\rm GC}, i}$, which depends on a set of model parameters which we call $\theta$.  We write this dependence explicitly as $p(L_{{\rm GC}, i}|\theta)$.
    \item Given the relation between the flux and luminosity, $\phi_i = L_{{\rm GC},i}/4\pi d_i^2$, where $d_i$ for each globular cluster is given in Table \ref{tab:tab7-9-hopper-linden}, it is possible to rewrite the luminosity function as function of the flux: $p_i(\phi_i|\theta)$. While the parameters $\theta$ would not depend on the specific globular cluster, the distance $d_i$ does. For this reason, given a gamma ray luminosity function, $p_i(\phi_i|\theta)$ will be different for each globular cluster.
    \item We combine the likelihood derived for a specific observation with the probability to observe a certain flux, and marginalize (integrate) over the flux in order to obtain the likelihood as a function of the parameter $\theta$: $\mathcal{L}_{\rm obs}^i(\theta)=\int_0^\infty \mathcal{L}_{\rm obs}(\phi_i)p_i(\phi_i|\theta)d\phi_i$
    \item We notice how each observation is independent from each other. For this reason, the joint likelihood on the entire set of globular clusters adopted in the analysis will be given by $\mathcal{L}_{\rm obs}(\theta)=\prod_i \mathcal{L}_{\rm obs}^i(\theta)$.
    \item Given the likelihood of the entire dataset $\mathcal{L}_{\rm obs}(\theta)$, it is now possible to derive a maximum likelihood estimation of the parameters $\theta$.
\end{enumerate}

\begin{figure}[t]
\centering
\includegraphics[width=0.49\textwidth]{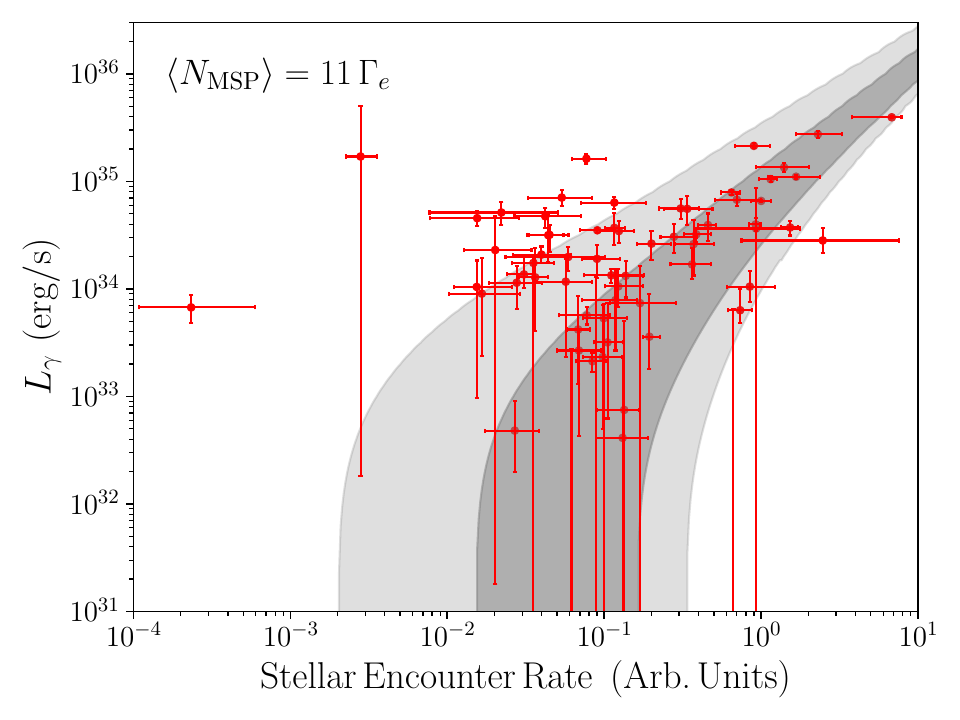}
\includegraphics[width=0.49\textwidth]{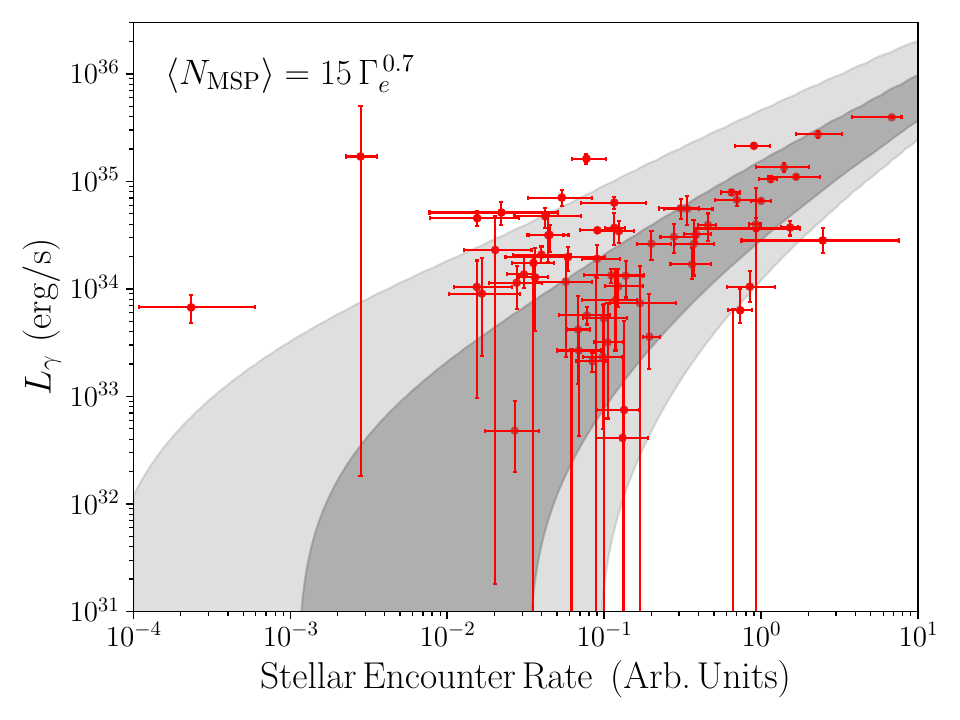}
\includegraphics[width=0.49\textwidth]{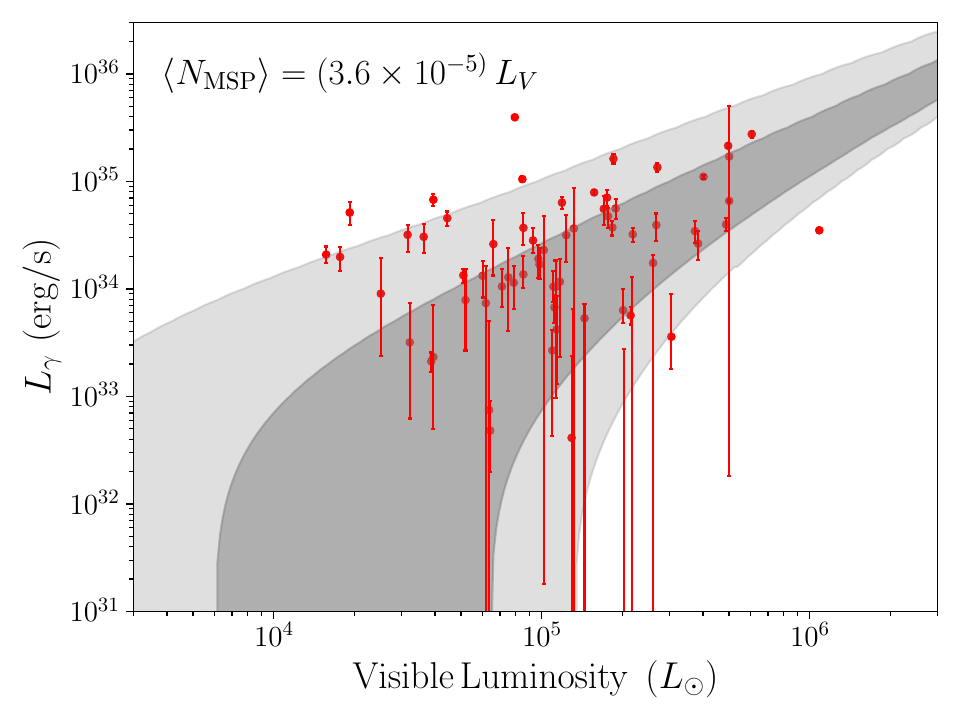}
\includegraphics[width=0.49\textwidth]{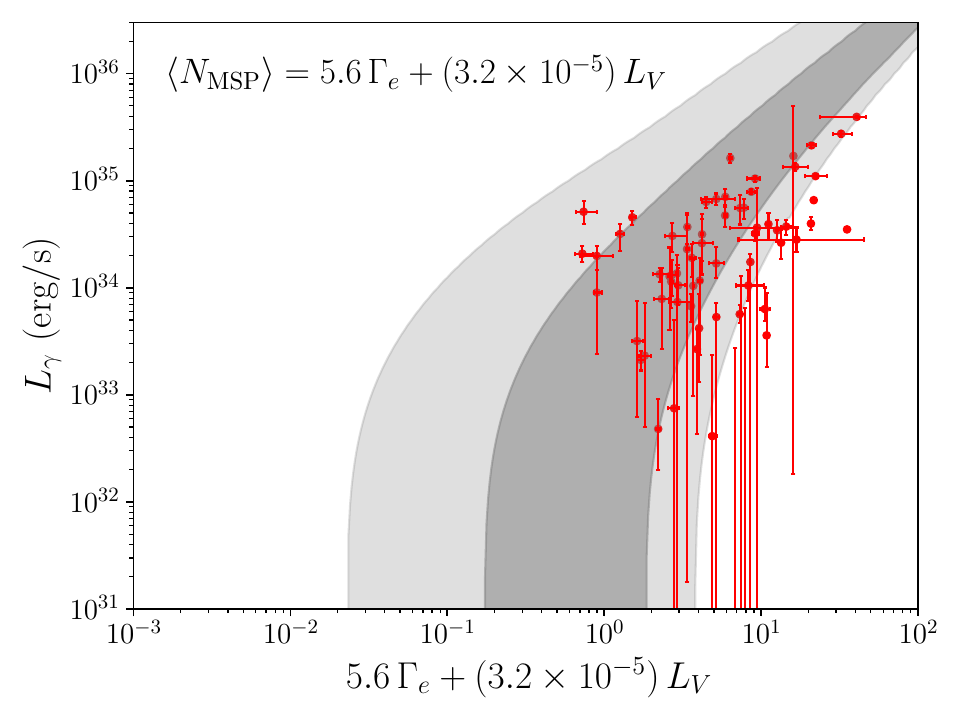}
\includegraphics[width=0.49\textwidth]{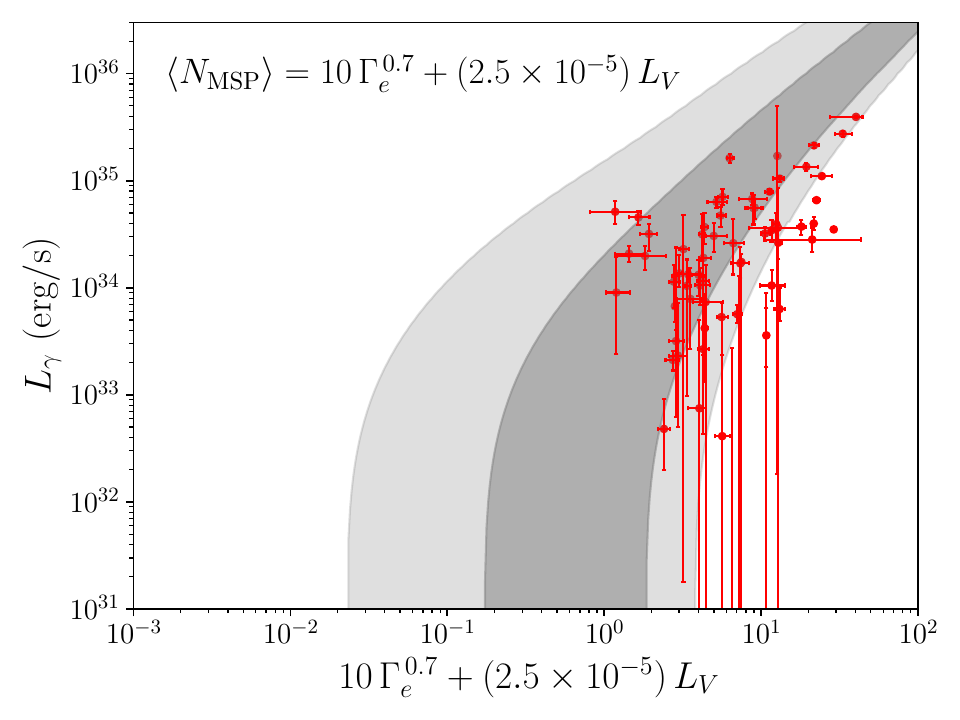}
\caption{The measured gamma-ray luminosities of the 87 globular clusters considered in this study (those with $L_V > 10^5 \, L_{\odot}$ or $\Gamma_e > 0.01$), compared to combinations of their stellar encounter rates and/or visible luminosities. As dark (light) grey bands, we show the 68.27\% (95.45\%) confidence interval predicted by the model for the the total gamma-ray luminosity. In each frame, we show the results that were obtained using the best-fit values for $\langle L_{\gamma} \rangle$, $\sigma_L$, $R_{\rm MSP}$, and/or $R'_{\rm MSP}$ (see Table~\ref{tab:param_constraints}).}
\label{fig:gamma1}
\end{figure}

In Fig.~\ref{fig:gamma1}, we show several examples of the results we obtain following this procedure. In each frame, we plot the measured gamma-ray luminosities of the 87 globular clusters considered in this portion of our study (those with $L_V > 10^5 \, L_{\odot}$ or $\Gamma_e > 0.01$), compared to their stellar encounter rates, visible luminosities, or selected combinations thereof. Then, as dark (light) grey bands, we plot the 68.27\% (95.45\%) range of the total gamma-ray luminosity from a given cluster that is predicted by the model in question. In each frame, we show the results as found adopting the best-fit values for $\langle L_{\gamma} \rangle$, $\sigma_L$, $R_{\rm MSP}$, and/or $R'_{\rm MSP}$ (see Table~\ref{tab:param_constraints}).

\begin{figure}[t]
\centering
\includegraphics[width=0.49\textwidth]{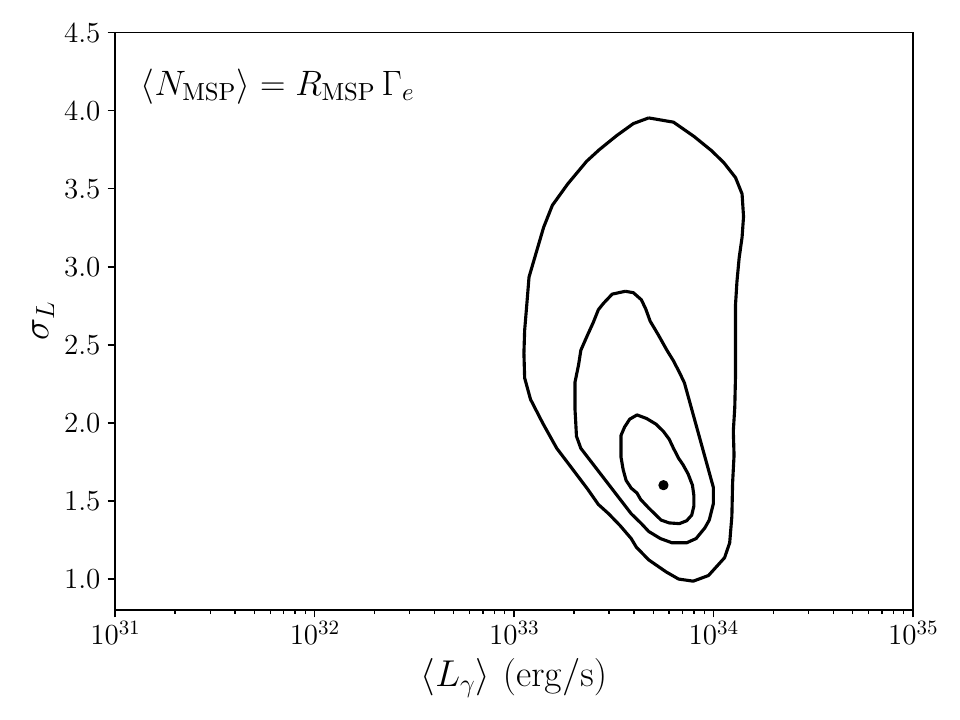}
\includegraphics[width=0.49\textwidth]{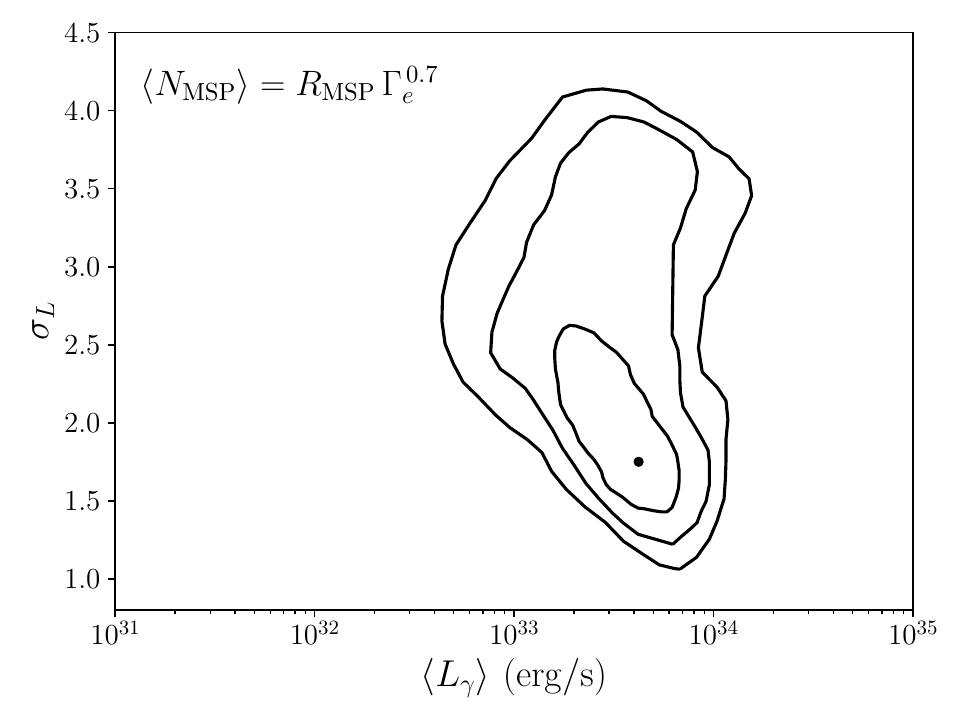}
\includegraphics[width=0.49\textwidth]{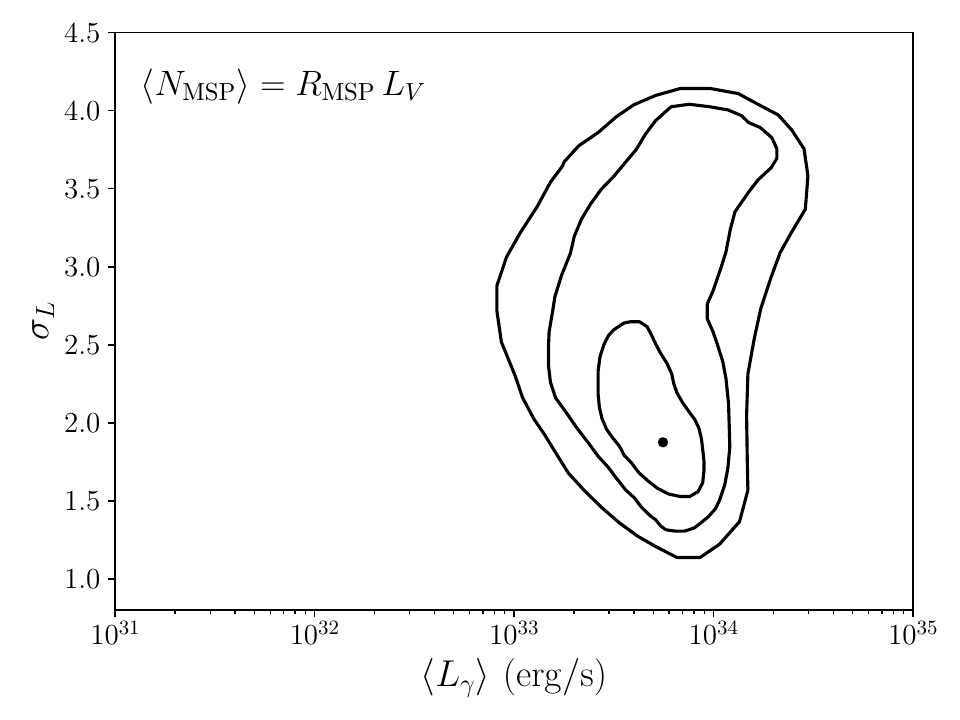}
\includegraphics[width=0.49\textwidth]{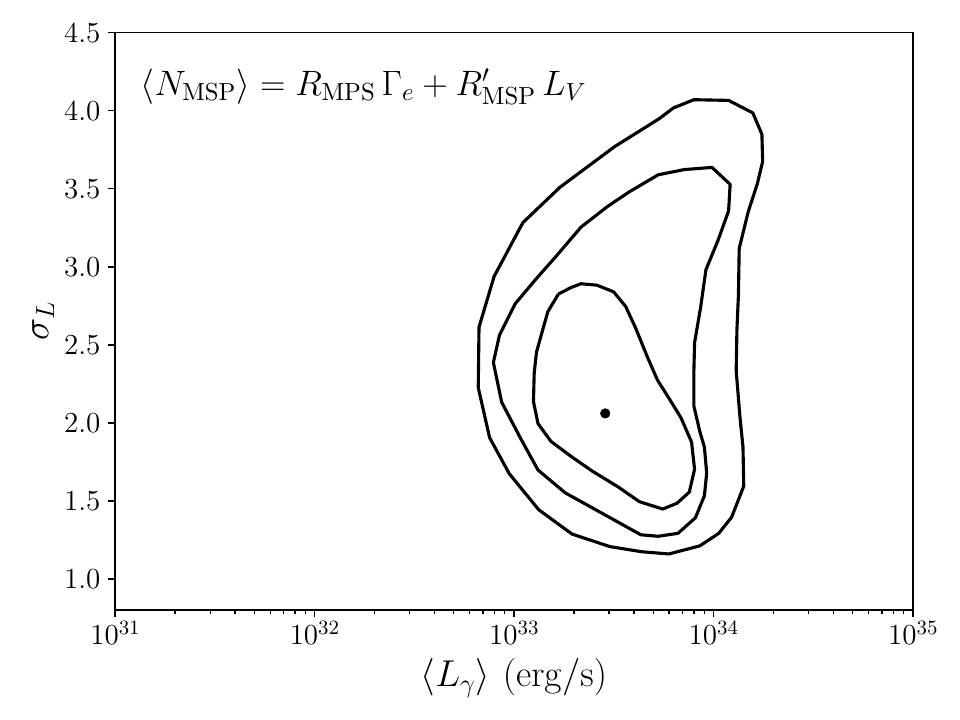}
\includegraphics[width=0.49\textwidth]{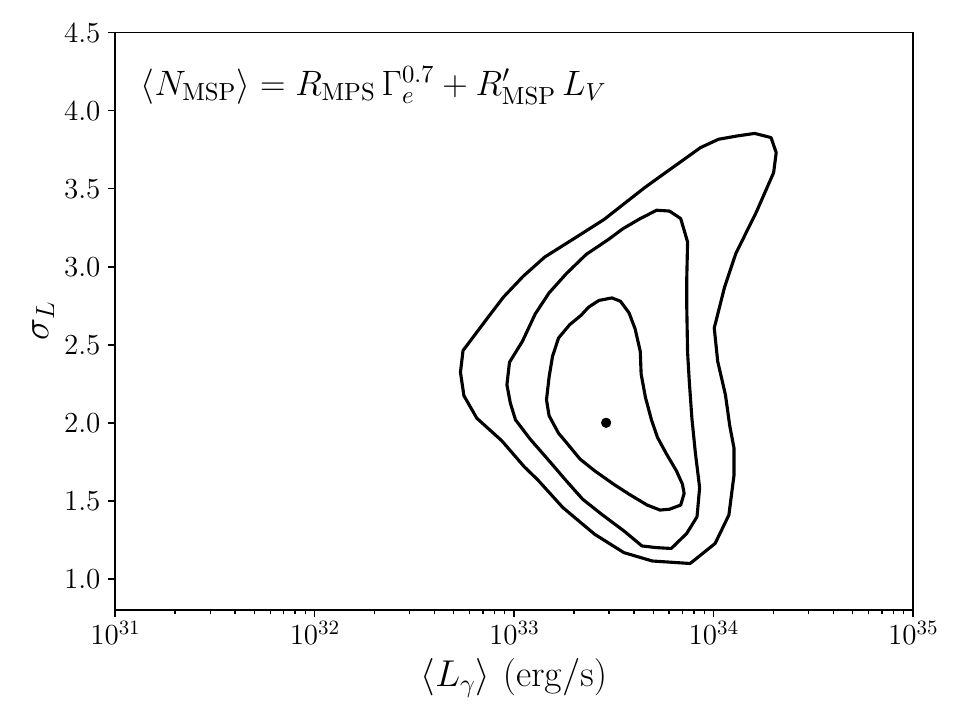}
\caption{Constraints on the MSP gamma-ray luminosity function parameters, after marginalizing over $R_{\rm MSP}$ and/or $R'_{\rm MSP}$. The best-fit point in each frame is shown as a dot, while the contours reflect the regions that are preferred at the levels of 1, 2 and 3$\sigma$ confidence.}
\label{fig:constraints}
\end{figure}

In Fig.~\ref{fig:constraints}, we plot the constraints on the luminosity function parameters, $\langle L_{\gamma} \rangle$ and $\sigma_L$, after marginalizing over the values of $R_{\rm MSP}$ and/or $R'_{\rm MSP}$. The best-fit point in each frame is shown by a dot, while the contours reflect the regions that are preferred at the levels of 1, 2 and 3$\sigma$ confidence. We show the results for each of the five hypotheses described in Sec.~\ref{sec:cases} for the relationship between the value of $\langle N_{\rm MSP}\rangle$ and those of $L_V$ and/or $\Gamma_e$, finding each case to yield qualitatively similar results. Namely, this procedure consistently favors scenarios with $\langle L_{\gamma}\rangle \sim (1-8)\times 10^{33}\, {\rm erg/s}$ and $\sigma_L \sim 1.4-2.7$. In Table~\ref{tab:param_constraints}, we show the values of the parameters favored by our analysis, including those for $R_{\rm MSP}$ and $R'_{\rm MSP}$. These parameters are directly related to the total number of gamma-ray MSPs contained within this sample of 87 globular clusters:  $N^{\rm tot}_{\rm MSP}=42.4 \, R_{\rm MSP} + 1.55 \times 10^7 \, R'_{\rm MSP}$ (for cases with $\langle N^{\rm tot}_{\rm MSP} \rangle \propto \Gamma_e$), or $N_{\rm MSP}=41.7 \, R_{\rm MSP}^{0.7} + 155 \times 10^7 \, R'_{\rm MSP}$ (for cases with $\langle N_{\rm MSP} \rangle \propto \Gamma_e^{0.7}$). For the parameters favored by our analysis, these 87 globular clusters contain a total of $\sim 400-1500$ gamma-ray emitting MSPs.

We note that the different hypotheses we have considered for the relationship between the value of $\langle N_{\rm MSP}\rangle$ and those of $L_V$ and/or $\Gamma_e$ can yield significantly different values for the overall likelihood. In Table~\ref{tab:models_compare}, we show the difference between these resulting likelihoods (for the best-fit parameters in each case). Note that these hypotheses have different numbers of degrees-of-freedom, and this should be taken in to account when comparing their relative likelihoods.

\begin{table*}[t]
\centering
\normalsize
\vspace{12pt}
{\renewcommand{\arraystretch}{1.2}
\begin{tabular}{|c|c|cccc|}
    \hline $\langle N_{\rm MSP}\rangle$ & Parameter & Best Fit & $\bm{1 \sigma}$ Range & $\bm{2 \sigma}$ Range & $\bm{3 \sigma}$ Range\\
    \hline\hline
      & $\langle L_{\gamma} \rangle$ & $5.6 \times 10^{33}$& $(3.4 - 8.1) \times 10^{33}$& $(2.0 - 10) \times 10^{33}$& $(1.1 - 13) \times 10^{33}$\\
      $R_{\rm MSP} \, \Gamma_e$ & $\sigma_L$ & $1.6$& $1.4 - 2.0$& $1.2 - 2.8$& $1.0 - 3.9$ \\
    & $R_{\rm MSP}$ & $11$& $9.5 - 15$& $7.6 - 32$& $6.4 - 110$ \\
    \hline
         & $\langle L_{\gamma} \rangle$ & $4.2 \times 10^{33}$& $(1.8 - 6.6) \times 10^{33}$& $(0.8 - 9.2) \times 10^{33}$& $(0.4 - 15) \times 10^{33}$\\
     $R_{\rm MSP} \, \Gamma_e^{\,0.7}$ & $\sigma_L$ & $1.8$& $1.4 - 2.6$& $1.1 - 3.9$& $1.0 - 4.1$ \\
    & $R_{\rm MSP}$ & $15$& $11 - 36$& $8.5 - 130$& $7.1 - 220$ \\
    \hline
         & $\langle L_{\gamma} \rangle$ & $5.6 \times 10^{33}$& $(2.7-8.7) \times 10^{33}$& $(1.4 - 21) \times 10^{33}$& $(0.8 - 29) \times 10^{33}$\\
     $R_{\rm MSP} \, L_V$ & $\sigma_L$ & $1.9$& $1.5 - 2.7$& $1.3 - 4.0$& $1.1 - 4.1$ \\
    & $R_{\rm MSP}$ & $3.6\times 10^{-5}$& $(2.7-6.5)\times 10^{-5}$& $(1.8-22)\times 10^{-5}$& $(1.4-35)\times 10^{-5}$ \\
    \hline    
         & $\langle L_{\gamma} \rangle$ & $2.9 \times 10^{33}$& $(1.3 - 5.5) \times 10^{33}$& $(0.9 - 8.4) \times 10^{33}$& $(0.7 - 14) \times 10^{33}$\\
     $R_{\rm MSP} \, \Gamma_e$ & $\sigma_L$ & $2.1$& $1.7 - 2.6$& $1.3 - 3.5$& $1.2 - 4.0$ \\
    $+R'_{\rm MSP}\,L_V$& $R_{\rm MSP}$ & $5.6$& $3.2 - 14$& $1.8 - 21$& $1.2 - 39$ \\
    & $R'_{\rm MSP}$ & $3.2\times 10^{-5}$& $(1.8-8.0)\times 10^{-5}$& $(1.4-11)\times 10^{-5}$& $(1.1-18)\times 10^{-5}$ \\
    \hline   
         & $\langle L_{\gamma} \rangle$ & $2.9 \times 10^{33}$& $(1.4 - 7.2) \times 10^{33}$& $(0.9 - 8.7) \times 10^{33}$& $(0.5 - 21) \times 10^{33}$\\
     $R_{\rm MSP} \, \Gamma_e^{0.7}$ & $\sigma_L$ & $2.0$& $1.4 - 2.8$& $1.2 - 3.4$& $1.1 - 3.9$ \\
    $+R'_{\rm MSP}\,L_V$& $R_{\rm MSP}$ & $10$& $6.4-23$& $3.9 - 66$& $2.9 - 110$ \\
    & $R'_{\rm MSP}$ & $2.5\times 10^{-5}$& $(1.6-5.7)\times 10^{-5}$& $(1.1-8.0)\times 10^{-5}$& $(0.8-11)\times 10^{-5}$ \\
    \hline   
\end{tabular}
}
\caption{The parameter values favored by our analysis. Values for the average MSP gamma-ray luminosity, $\langle L_{\gamma} \rangle$, are given in erg/s.}
\label{tab:param_constraints}
\end{table*}

\begin{table*}[t]
\centering
\normalsize
\vspace{12pt}
{\renewcommand{\arraystretch}{1.2}
\begin{tabular}{|c|c|}
    \hline $\langle N_{\rm MSP}\rangle$ & $-2\Delta \ln \mathcal{L}$\\
    \hline\hline
       $R_{\rm MSP} \, \Gamma_e$ & 146.0 \\
    \hline
       $R_{\rm MSP} \, \Gamma_e^{\,0.7}$ & 45.4 \\
    \hline
       $R_{\rm MSP} \, L_V$ & 37.7 \\
    \hline    
       $R_{\rm MSP} \, \Gamma_e +R'_{\rm MSP}\,L_V$ & 13.5 \\
    \hline  
       $R_{\rm MSP} \, \Gamma_e^{\,0.7} +R'_{\rm MSP}\,L_V$ & 0.0 \\
    \hline      
\end{tabular}
}
\caption{The change in twice the log-likelihood of our analysis, relative to the best-fit case shown in the bottom row of this table.}
\label{tab:models_compare}
\end{table*}

So far, we have assumed that the mean number of MSPs in a given globular cluster is determined entirely by its luminosity and/or stellar encounter rate, without any departures from this prediction (other than the scatter inherent in each cluster's draw from the Poisson distribution for $N_{\rm MSP}$). In reality, our models for determining $\langle N_{\rm MSP} \rangle$ cannot possibly contain all of the relevant information and, therefore, should not be perfectly predictive. This leads us to expect that there should some degree of variation in the mean number of MSPs in given cluster, beyond that which can be predicted by the values of $L_V$ and $\Gamma_e$ alone. 

To assess the impact of such cluster-to-cluster variation, we take the mean number of expected MSPs to have an intrinsic degree of scatter, which we model according to a log-normal distribution with a width, $\sigma_N$. This more complex model represents a potentially more conservative approach, and allows us to accounting for potential unmodeled systematic uncertainties. By rerunning our analysis, we find that in the case of our best fit model, $\langle N_{\rm MSP}\rangle = R_{\rm MSP} \, \Gamma_e^{0.7} +R'_{\rm MSP} \, L_V$, the overall fit is improved by a maximum of $-2\Delta \ln \mathcal{L} \approx 8$ for $\sigma_N \approx 0.5$. In this new best fit, the values of $\langle L_{\gamma} \rangle$ and $\sigma_L$ are shifted modestly from the case without this intrinsic scatter, from $\langle L_{\gamma} \rangle \approx 2.9 \times 10^{33} \, {\rm erg/s}$ and $\sigma_L \approx 2.0$ to $\langle L_{\gamma} \rangle \approx 1.8\times 10^{33} \, {\rm erg/s}$ and $\sigma_L \approx 1.4$.

While an improvement of $-2\Delta \ln \mathcal{L} \approx 8$ corresponds to a statistical preference between 2$\sigma$ and 3$\sigma$, this more complex model also introduces new modeling uncertainties. Given this, we cannot exclude that this improvement is due to statistical fluctuations or unknown systematic uncertainties. We therefore refrain from adopting the model with cluster-to-cluster variations as our fiducial baseline, and instead opt for the simpler fitting formula without variations, which has a stronger basis in the literature.

Before moving on, we will use these results to comment on the previous studies that have interpreted the gamma-ray emission observed from NGC 5139~\cite{Reynoso-Cordova:2019biv} and NGC 104~\cite{Brown:2018pwq} to be possible signals of annihilating dark matter. Although this interpretation cannot be easily ruled out, our results in Fig.~\ref{fig:gamma1} demonstrate why no such exotic emission is required. In fact, these clusters have values of $\Gamma_e$ and $L_V$ that are entirely compatible with their observed gamma-ray luminosities. No dark matter annihilation or other exotic contributions are required.

\section{Comparisons with Previous Studies}

\begin{figure}[t]
\centering
\includegraphics[width=0.66\textwidth]{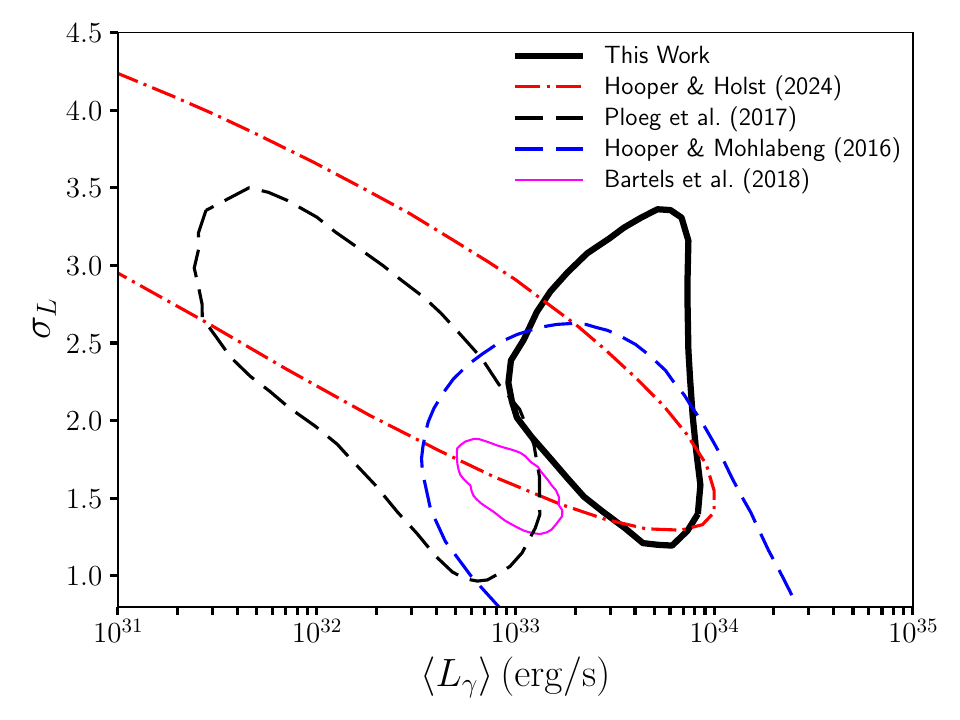}
\caption{A comparison of the $2\sigma$ constraints on the MSP gamma-ray luminosity found in this study (for the best-fitting case of $\langle N_{\rm MSP}\rangle = R_{\rm MSP} \, \Gamma_e^{0.7} +R'_{\rm MSP} \, L_V$) to those published in previous studies of MSPs in the Galactic Plane. The luminosity function preferred by our analysis is in good agreement with those derived by Holst and Hooper~\cite{Holst:2024fvb}, and by Hooper and Mohlabeng~\cite{Hooper:2016rap}, but favor somewhat larger values for the average MSP luminosity than Ploeg {\it et al.}~\cite{Ploeg:2017vai}, or Bartels {\it et al.}~\cite{Bartels:2018xom}.}
\label{fig:compare}
\end{figure}

A number of previous studies have attempted to measure or constrain the gamma-ray luminosity function of MSPs~\cite{Holst:2024fvb,Cholis:2014noa,Hooper:2016rap,Hooper:2015jlu,Bartels:2018xom,Ploeg:2017vai,Ploeg:2020jeh}). In Fig.~\ref{fig:compare}, we compare the results of this study (for the best-fitting case of $\langle N_{\rm MSP}\rangle = R_{\rm MSP} \, \Gamma_e^{0.7} +R'_{\rm MSP} \, L_V$) with those published in previous studies of the MSP population within the Galactic Plane. The luminosity function preferred by our analysis is in good agreement with those recently derived by Holst and Hooper~\cite{Holst:2024fvb},  and previously by Hooper and Mohlabeng~\cite{Hooper:2016rap}. Our results, however, favor somewhat larger values of the average MSP luminosity, $\langle L_{\gamma}\rangle$, than the analyses of Ploeg {\it et al.}~\cite{Ploeg:2017vai} or Bartels {\it et al.}~\cite{Bartels:2018xom}.

We note that the analysis of Bartels {\it et al.}~\cite{Bartels:2018xom} made use of pulsar distance measurements, which are
often subject to significant and difficult to quantify systematic uncertainties. For most pulsars, distance estimates are derived
from measurements of their dispersion measure (ie., the frequency-dependent time delay of the radio pulses, which can be used to determine the column density of free electrons along the pulsar’s line-of-sight). When the dispersion measure of a pulsar is combined with a model of the free electron distribution, one can obtain an estimate of its distance. Such estimates are known to vary significantly, however, depending on the electron distribution model that is adopted~\cite{Yao:2017kcp,Cordes:2002wz}, and sometimes disagree with more reliable parallax determinations by factors of up to $\sim 3$~\cite{Fermi-LAT:2023zzt}.

It is possible that the MSPs population found within globular clusters could feature a gamma-ray luminosity function that is different from those pulsars in the Galactic Plane. This could be motivated by the fact that globular clusters are known to contain an old stellar population, and their pulsars could plausibly have already lost much of rotational kinetic energy, compared to that of more recently formed MSPs. This difference, however, would lead us to expect MSPs in globular clusters to be systematically {\it fainter} than those in the Galactic Plane, which is opposite to that found in the studies of Ploeg {\it et al.}~\cite{Ploeg:2017vai} and Bartels {\it et al.}~\cite{Bartels:2018xom}.   

We can more directly compare our results to those of the 2016 analysis by Hooper and Linden~\cite{Hooper:2016rap}, which also studied MSPs in globular clusters, using in their case 7 years of Fermi data (compared to the 15.8 years of data used in this analysis). The 2016 study favored a MSP gamma-ray luminosity function with $\langle L_{\gamma} \rangle \sim (0.7-5) \times 10^{34} \, {\rm erg/s}$ and $\sigma_L \sim 1-3$. While the results presented here are statistically consistent with those of that study (at the approximately $\sim 2\sigma$ level), our analysis tends to prefer somewhat smaller values of $\langle L_{\gamma} \rangle$ and larger values of $R_{\rm MSP}$, by a factor of $\sim 2-8$. For a more detailed comparison, see Appendix \ref{sec:app-comparison}.

\section{Implications for the Galactic Center Gamma-Ray Excess}

As stated in the introduction, very few pulsars have been detected in the vicinity of the Inner Galaxy. If the GCE is generated by MSPs, this fact requires the MSPs responsible for this signal to be faint enough to have gone undetected. This, in turn, would require the MSPs in the Inner Galaxy to be systematically lower in luminosity than those found in other environments~\cite{Holst:2024fvb} (see also, Refs.~\cite{Cholis:2014noa,Hooper:2016rap,Hooper:2015jlu,Bartels:2018xom,Ploeg:2017vai,Ploeg:2020jeh,Dinsmore:2021nip}).

\begin{figure}[t]
\centering
\includegraphics[width=0.66\textwidth]{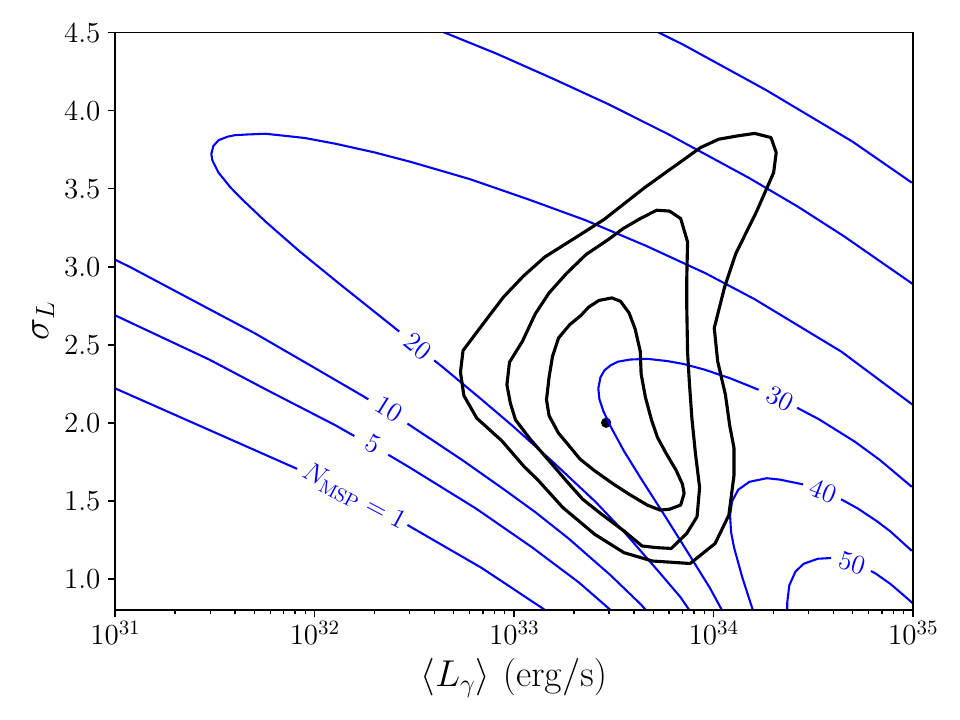}
\caption{The constraints on the MSP gamma-ray luminosity found in this study (shown for the case of $\langle N_{\rm MSP}\rangle = R_{\rm MSP}\, \Gamma_e^{0.7} +R'_{\rm MSP} \,L_V$), shown along with contours representing the number of Inner Galaxy MSPs that should
appear in the Fermi source catalogs, $N_{\rm MSP}$, assuming that the Galactic Center Gamma-Ray Excess (GCE) is generated by MSPs~\cite{Holst:2024fvb}. As there are only three Inner Galaxy pulsars candidates in the 3PC, MSPs can generate the excess
gamma-ray emission only if that source population consists of members that are significantly less luminous than those present in globular clusters.}
\label{fig:3PC}
\end{figure}

The possibility that the Inner Galaxy could contain a large population of low-luminosity pulsars has been motivated by the large proportion of old stars that are found in the Galactic Bulge relative to those in the Galactic Plane. The stellar populations found in globular clusters, however, are comparably old to those in the Bulge. By measuring the gamma-ray luminosity function of MSPs in globular clusters, we can hope to shed some light on the question of whether the Inner Galaxy might contain a very large abundance of very low-luminosity MSPs. 

In Fig.~\ref{fig:3PC}, we plot the number of MSPs that should have been detected by Fermi and contained within their source catalogs~\cite{Fermi-LAT:2023zzt} if they were responsible for producing the GCE, as a function of the luminosity function parameters~\cite{Holst:2024fvb}. If the GCE is generated by MSPs with the same luminosity function and other characteristics as those found in globular clusters, Fermi's source catalogs would have contained $N_{\rm MSP}\sim 17-37$ individual members of this population. Given that only three millisecond pulsars have been detected with a direction and distance that does not preclude it from being part of an Inner Galaxy population, the luminosity function derived here is incompatible with pulsar interpretations of the GCE, unless a large number of unassociated Fermi sources are, in fact, unidentified MSPs. Given systematic efforts that have been conducted to search for radio pulsations from unassociated Fermi sources, this possibility seems unlikely. Alternatively, the observed gamma-ray excess could be generated by MSPs if those pulsars are significantly less luminous, on average, than those located in globular clusters or in the Galactic Plane. 

\begin{figure}[t]
\centering
\includegraphics[width=0.66\textwidth]{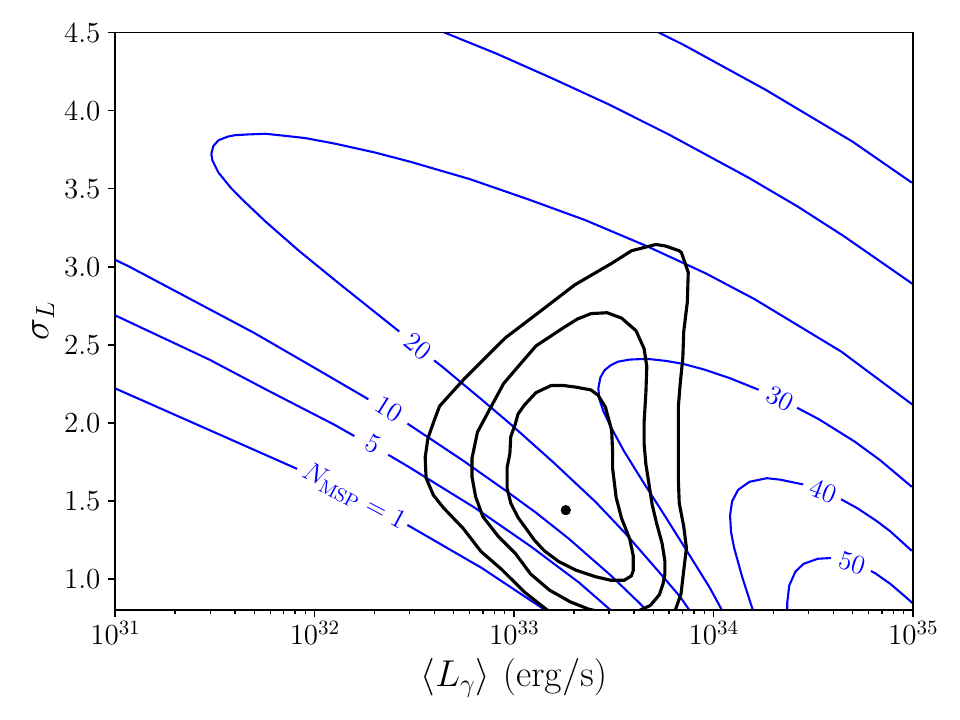}
\caption{ As in Figure \ref{fig:3PC}, but for the best fit model including cluster-to-cluster variations. The results are compatible with the estimates in Figure \ref{fig:3PC} within the $1\sigma$ band, although the model with cluster-to-cluster variations favors lower values for $\langle L_{\gamma} \rangle$ and $\sigma_L$.}
\label{fig:3PC-c2c-var}
\end{figure}

As mentioned in Section \ref{sec:glf}, our models for the mean number of millisecond pulsars in a given globular cluster are subject to intrinsic uncertainties. To account for this, we also considered a more flexible model that includes an intrinsic cluster-to-cluster variation in the expected number of MSPs. The purpose of this approach is to derive a more conservative lower bound on the number of pulsars that should have been detected in the Inner Galaxy if they are responsible for the Galactic Center Excess. In Figure \ref{fig:3PC-c2c-var}, we show the results of this analysis. By allowing for a wider range of pulsar populations, particularly those that are fainter on average, this model lowers the number of expected detections to the range of $N_{\rm MSP} \approx 3 -30$.  Although this bound is less constraining by construction, the results nonetheless remain in tension at a level of at least $2\sigma$ with the three identified pulsar candidates.

\section{Summary and Conclusions}

In this study, we have analyzed the gamma-ray emission from the Milky Way's globular clusters, using 15.8 years of publicly available data collected by the Fermi Gamma-Ray Space Telescope. We report the robust detection of 56 globular clusters with a statistically significant test statistic of TS > 25, 8 of which are not contained in previous gamma-ray source catalogs. 

Making use of previously established correlations between the mean number of millisecond pulsars in a globular cluster and the stellar encounter rate or visible luminosity of that cluster, we perform a determination of the gamma-ray luminosity function of the millisecond pulsars in the Milky Way's globular cluster system. To this end, we have considered 87 globular clusters with high stellar encounter rates and/or visible luminosities, and combined this information with our measurements of their gamma-ray luminosities. We then fitted the data to a log-normal luminosity function, finding it to favor a mean pulsar luminosity of $\langle L_{\gamma}\rangle \sim (1-8)\times 10^{33}\, {\rm erg/s}$ (integrated between 0.1 and 100 GeV), and a width of $\sigma_L \sim 1.4-2.7$. Our analysis also indicates that the Milky Way's globular cluster system contains a total of $\sim 400-1500$ gamma-ray emitting millisecond pulsars. The luminosity function favored by this analysis is consistent with that exhibited by millisecond pulsars found in the Galactic Plane~\cite{Holst:2024fvb}.  

The leading astrophysical interpretation of the Galactic Center Gamma-Ray Excess has long been that this signal could be generated by large population of centrally-located millisecond pulsars. Very few pulsars have been detected in the Inner Galaxy, however, forcing us to place strong constraints on the gamma-ray luminosity function of any pulsar population that could be responsible for the excess emission. In particular, such a pulsar population would have to be systematically less bright than those found in the Galactic Plane~\cite{Holst:2024fvb}. 

It has been suggested that the Inner Galaxy pulsar population could be older and therefore less luminous, on average, than pulsars found in the Galactic Plane. The stars in globular clusters are also very old, however. Thus if the Inner Galaxy pulsar population is very faint, one would expect the same to be true among the pulsars found in globular clusters. The results of this study indicate that this is not the case. This result poses a challenge to pulsar interpretations of the Galactic Center Gamma-Ray Excess and, by extension, potentially bolsters dark matter interpretations of this long-standing signal. 

\bigskip

\acknowledgments
AA acknowledges support from Generalitat Valenciana through the “GenT program”, ref.: CIDEGENT/2020/055 (PI: B. Zaldivar). DH is supported by the Office of the Vice Chancellor for Research at the University of Wisconsin–Madison with funding from the Wisconsin Alumni Research Foundation. TL acknowledges support by the Swedish Research Council under contract 2022-04283. The authors also acknowledge the computational resources at Artemisa and the technical support provided by the Instituto de Fisica Corpuscular, IFIC(CSIC-UV). Artemisa is co-funded by the European Union through the 2014-2020 ERDF Operative Programme of Comunitat Valenciana, project IDIFEDER/2018/048.

\bibliography{bibliography}

\appendix
\section{Additional Tables}
\label{sec:app-tables}
In this section, we provide several tables which contain information regarding each of the globular clusters considered in our study, as well as the results of our gamma-ray analysis for each of these sources.

\begingroup
\renewcommand\arraystretch{1.3}

\begin{longtable}{|l|l|l|c|c|c|} 
\multicolumn{6}{c}{Table \thetable} \\ 
\hline
Globular Cluster & Alternate Name & {Flux (erg/cm$^2$/s)} & $\alpha$ & {$E_{c}$ (GeV)} & TS \\
\hline
\hline
\endfirsthead %

\multicolumn{6}{c}{Table \thetable\ -- continued from previous page} \\ 
\hline
Globular Cluster & Alternate Name & {Flux (erg/cm$^2$/s)} & $\alpha$ & {$E_{c}$ (GeV)} & TS \\
\hline
\hline
\endhead %

\multicolumn{6}{r}{continued on next page} \\ 
\endfoot %

\hline
\noalign{\vspace{10pt}}
\captionsetup{justification=justified} %
\caption{The gamma-ray flux (integrated between $0.1-100$ GeV) and the best-fit spectral parameters for each of the 56 globular clusters that were detected in our analysis with TS > 25.}
\endlastfoot %
NGC 104 & 47 Tuc & $2.773^{+0.041}_{-0.051} \times 10^{-11}$ & 1.37 & 2.85 & 10801.3 \\\hline
Terzan 5 & Terzan 11 & $9.245^{+0.215}_{-0.181} \times 10^{-11}$ & 1.75 & 4.24 & 5955.9 \\\hline
NGC 6266 & M62 & $1.984^{+0.091}_{-0.097} \times 10^{-11}$ & 1.44 & 2.79 & 1504.2 \\\hline
NGC 6388 & & $1.823^{+0.091}_{-0.090} \times 10^{-11}$ & 1.41 & 2.31 & 1399.8 \\\hline
NGC 6626 & M28 & $2.170^{+0.111}_{-0.118} \times 10^{-11}$ & 1.52 & 2.35 & 1346.0 \\\hline
NGC 5139 & Omega Centauri & $1.100^{+0.056}_{-0.054} \times 10^{-11}$ & 1.21 & 2.11 & 1168.9 \\\hline
NGC 6624 & & $1.405^{+0.086}_{-0.090} \times 10^{-11}$ & 1.60 & 4.75 & 902.9 \\\hline
GLIMPSE01 & & $4.305^{+0.256}_{-0.238} \times 10^{-11}$ & 1.55 & 4.22 & 686.5 \\\hline
NGC 6441 & & $1.710^{+0.112}_{-0.120} \times 10^{-11}$ & 1.85 & 3.91 & 675.9 \\\hline
NGC 6440 & & $1.586^{+0.141}_{-0.142} \times 10^{-11}$ & 1.90 & 6.41 & 399.2 \\\hline
NGC 6752 & & $3.954^{+0.444}_{-0.419} \times 10^{-12}$ & 1.07 & 1.46 & 341.5 \\\hline
NGC 6316 & & $1.248^{+0.115}_{-0.119} \times 10^{-11}$ & 1.86 & 5.12 & 336.2 \\\hline
GLIMPSE02 & & $2.568^{+0.233}_{-0.238} \times 10^{-11}$ & 1.78 & 2.81 & 203.7 \\\hline
NGC 6652 & & $5.664^{+0.694}_{-0.676} \times 10^{-12}$ & 1.87 & 4.78 & 191.8 \\\hline
2MS-GC01 & 2MASS-GC01 & $2.666^{+0.226}_{-0.238} \times 10^{-11}$ & 1.66 & 1.86 & 191.8 \\\hline
NGC 2808 & & $3.625^{+0.489}_{-0.484} \times 10^{-12}$ & 1.69 & 4.06 & 182.1 \\\hline
NGC 6541 & & $4.747^{+0.666}_{-0.670} \times 10^{-12}$ & 1.85 & 4.80 & 154.8 \\\hline
NGC 6093 & M80 & $4.402^{+0.857}_{-0.807} \times 10^{-12}$ & 1.67 & 6.78 & 146.0 \\\hline
NGC 6569 & & $4.984^{+0.821}_{-0.779} \times 10^{-12}$ & 1.62 & 6.64 & 133.8 \\\hline
NGC 7078 & & $3.608^{+0.488}_{-0.517} \times 10^{-12}$ & 2.04 & 2.19 & 127.4 \\\hline
NGC 6717 & Palomar 9& $3.454^{+0.619}_{-0.560} \times 10^{-12}$ & 1.29 & 3.26 & 121.2 \\\hline
NGC 6656 & M22 & $4.557^{+0.882}_{-0.784} \times 10^{-12}$ & 0.82 & 1.00 & 120.5 \\\hline
NGC 1904 & M79 & $3.176^{+0.373}_{-0.362} \times 10^{-12}$ & 2.81 & 56.20 & 100.9 \\\hline
Palomar 6 & & $1.137^{+0.173}_{-0.162} \times 10^{-11}$ & 2.20 & $\geq$ 100 & 94.9 \\\hline
Terzan 1 & HP 2 & $6.774^{+0.862}_{-0.871} \times 10^{-12}$ & 0.00 & 1.61 & 91.0 \\\hline
NGC 6139 & & $4.583^{+0.957}_{-0.911} \times 10^{-12}$ & 1.64 & 3.65 & 84.5 \\\hline
NGC 6254 & M10 & $2.295^{+0.407}_{-0.332} \times 10^{-12}$ & 0.00 & 0.92 & 79.7 \\\hline
Terzan 2 & HP 3 & $7.670^{+1.844}_{-1.672} \times 10^{-12}$ & 1.97 & 17.66 & 76.4 \\\hline
NGC 1851 & & $2.144^{+0.312}_{-0.314} \times 10^{-12}$ & 2.45 & $\geq$ 100 & 76.2 \\\hline
NGC 6544 & & $1.280^{+0.184}_{-0.193} \times 10^{-11}$ & 1.46 & 0.65 & 76.2 \\\hline
NGC 6341 & M92 & $1.334^{+0.340}_{-0.307} \times 10^{-12}$ & 1.38 & 1.95 & 73.7 \\\hline
NGC 6402 & M14 & $3.379^{+0.775}_{-0.706} \times 10^{-12}$ & 1.52 & 2.54 & 68.0 \\\hline
Terzan 9 & & $5.727^{+1.324}_{-0.947} \times 10^{-12}$ & 0.00 & 0.58 & 67.5 \\\hline
NGC 6397 & & $3.363^{+0.671}_{-0.653} \times 10^{-12}$ & 2.08 & 5.21 & 67.3 \\\hline
NGC 5904 & M5 & $1.552^{+0.432}_{-0.374} \times 10^{-12}$ & 1.15 & 2.30 & 66.8 \\\hline
Terzan 6 & HP 5 & $5.145^{+1.434}_{-1.163} \times 10^{-12}$ & 0.38 & 1.21 & 59.4 \\\hline
NGC 6273 & M19 & $2.853^{+0.837}_{-0.808} \times 10^{-12}$ & 1.88 & 21.03 & 53.1 \\\hline
Whiting 1 & & $1.156^{+0.298}_{-0.277} \times 10^{-12}$ & 1.37 & 15.60 & 52.4 \\\hline
Liller 1 & & $9.263^{+2.229}_{-2.224} \times 10^{-12}$ & 1.63 & 5.60 & 51.1 \\\hline
NGC 6528 & & $4.068^{+1.252}_{-1.128} \times 10^{-12}$ & 1.08 & 1.20 & 47.2 \\\hline
UKS 1 & & $1.068^{+0.199}_{-0.193} \times 10^{-11}$ & 0.23 & 0.25 & 45.6 \\\hline
NGC 362 & & $7.162^{+3.977}_{-1.624} \times 10^{-13}$ & 0.35 & 1.27 & 44.6 \\\hline
NGC 6637 & M 69 & $2.061^{+0.678}_{-0.675} \times 10^{-12}$ & 1.45 & 5.22 & 43.8 \\\hline
NGC 6723 & & $2.529^{+0.623}_{-0.612} \times 10^{-12}$ & 1.75 & 5.50 & 43.6 \\\hline
NGC 6218 & M12 & $3.024^{+0.657}_{-0.730} \times 10^{-12}$ & 2.39 & $\geq$ 100 & 42.9 \\\hline
NGC 6539 & & $6.541^{+1.290}_{-1.355} \times 10^{-12}$ & 2.66 & $\geq$ 100 & 39.0 \\\hline
NGC 5286 & & $2.418^{+0.660}_{-0.657} \times 10^{-12}$ & 2.23 & 18.15 & 35.8 \\\hline
NGC 6304 & & $2.557^{+1.101}_{-0.853} \times 10^{-12}$ & 0.92 & 1.69 & 34.8 \\\hline
NGC 6342 & & $3.671^{+0.874}_{-1.082} \times 10^{-12}$ & 2.32 & $\geq$ 100 & 33.1 \\\hline
NGC 6712 & & $2.381^{+1.123}_{-0.561} \times 10^{-12}$ & 0.00 & 0.67 & 31.4 \\\hline
NGC 6380 & Ton 1 & $2.617^{+0.886}_{-0.773} \times 10^{-12}$ & 0.75 & 1.77 & 29.9 \\\hline
NGC 6517 & & $4.119^{+1.208}_{-1.183} \times 10^{-12}$ & 1.76 & 2.06 & 29.2 \\\hline
IC 1257 & & $1.529^{+0.657}_{-0.468} \times 10^{-12}$ & 0.58 & 2.33 & 29.0 \\\hline
NGC 6522 & & $2.389^{+0.968}_{-0.588} \times 10^{-12}$ & 0.00 & 0.75 & 28.1 \\\hline
NGC 6540 & Djorg 3 & $2.841^{+1.163}_{-0.937} \times 10^{-12}$ & 1.16 & 1.95 & 28.1 \\\hline
NGC 6352 & & $2.532^{+0.596}_{-0.635} \times 10^{-12}$ & 0.00 & 0.25 & 27.9

\label{tab:high_TS}
\end{longtable}
\endgroup

\newpage
\begingroup
\renewcommand\arraystretch{1.3}
\begin{longtable}{|l |l| l| c|} 
\hline
Globular Cluster & Alternate Name & {Flux (erg/cm$^2$/s)} & TS \\
\hline
\hline
\endfirsthead %

\multicolumn{4}{c}{Table \thetable\ -- continued from previous page} \\ 
\hline
Globular Cluster & Alternate Name & {Flux (erg/cm$^2$/s)} & TS \\
\hline
\hline
\endhead %

\multicolumn{4}{r}{continued on next page} \\ 
\endfoot %

\hline
\noalign{\vspace{10pt}}
\captionsetup{justification=justified} %
\caption{The $2\sigma$ upper limits on the gamma-ray flux (integrated between $0.1-100$ GeV) for globular clusters that were detected in our analysis with TS > 25, but with spectra that peak at energies below 0.4 GeV.}
\endlastfoot %
NGC 6838 & M 71 & $<5.65 \times 10^{-12}$ & 62.84 \\\hline
Lynga 7 & BH184 & $<7.18 \times 10^{-12}$ & 54.87 \\\hline
AM 1 & E 1& $<1.47 \times 10^{-12}$ & 36.99 \\\hline
AM 4 & & $<2.88 \times 10^{-12}$ & 34.07 \\\hline
Palomar 1 & & $<2.00 \times 10^{-12}$ & 31.94 \\\hline
NGC 6809 & M 55 & $<3.42 \times 10^{-12}$ & 27.05 \\\hline
Palomar 10 & & $<6.40 \times 10^{-12}$ & 27.04

\label{tab:low_peak_TS}
\end{longtable}
\endgroup

\begingroup
\renewcommand\arraystretch{1.3}
\begin{longtable}{|l|l|l|c|} 
\multicolumn{4}{c}{Table \thetable} \\ 
\hline
Globular Cluster & Alternate Name & {Flux (erg/cm$^2$/s)} & TS \\
\hline
\hline
\endfirsthead %

\multicolumn{4}{c}{Table \thetable\ -- continued from previous page} \\ 
\hline
Globular Cluster & Alternate Name & {Flux (erg/cm$^2$/s)} & TS \\
\hline
\hline
\endhead %

\multicolumn{4}{r}{continued on next page} \\ 
\endfoot %

\hline
\noalign{\vspace{10pt}}
\captionsetup{justification=justified} %
\caption{The $2\sigma$ upper limits on the gamma-ray flux (integrated between $0.1-100$ GeV) for those globular clusters that were not significantly detected in our analysis (TS < 25).}
\endlastfoot %
IC 4499 & & $<1.83\times 10^{-12}$ & 15.53 \\\hline
E 3 & & $<2.31\times 10^{-12}$ & 23.25 \\\hline
NGC 4372 & & $<2.35\times 10^{-12}$ & 20.94 \\\hline
NGC 6101 & & $<1.88\times 10^{-12}$ & 21.99 \\\hline
NGC 4833 & & $<3.88\times 10^{-12}$ & 0.00 \\\hline
NGC 6362 & & $<2.52\times 10^{-12}$ & 0.00 \\\hline
NGC 1261 & & $<1.97\times 10^{-12}$ & 0.00 \\\hline
NGC 6584 & & $<1.04\times 10^{-12}$ & 4.92 \\\hline
Rup 106 & & $<3.80\times 10^{-12}$ & 0.90 \\\hline
NGC 5927 & & $<1.57\times 10^{-12}$ & 3.78 \\\hline
NGC 5946 & & $<4.57\times 10^{-12}$ & 0.00 \\\hline
BH 176 & & $<2.92\times 10^{-12}$ & 7.08 \\\hline
FSR 1735 & & $<1.35\times 10^{-11}$ & 0.00 \\\hline
ESO-SC06 & ESO280-SC06& $<5.01\times 10^{-12}$ & 0.00 \\\hline
NGC 3201 & & $<6.69\times 10^{-13}$ & 1.51 \\\hline
NGC 6496 & & $<4.16\times 10^{-12}$ & 0.00 \\\hline
Ton 2 & Pismis 26& $<3.39\times 10^{-12}$ & 18.01 \\\hline
NGC 5986 & & $<2.77\times 10^{-12}$ & 0.00 \\\hline
Pyxis & & $<3.64\times 10^{-12}$ & 1.86 \\\hline
NGC 6256 & & $<7.57\times 10^{-12}$ & 0.00 \\\hline
NGC 2298 & & $<2.72\times 10^{-12}$ & 0.14 \\\hline
Terzan 3 & & $<5.80\times 10^{-12}$ & 0.06 \\\hline
Terzan 7 & & $<3.59\times 10^{-12}$ & 0.00 \\\hline
NGC 6453 & & $<3.23\times 10^{-12}$ & 10.92 \\\hline
Terzan 8 & & $<4.20\times 10^{-12}$ & 2.46 \\ \hline
NGC 5824 & & $<7.52\times 10^{-13}$ & 2.91 \\\hline
Djorg 1 & & $<1.45\times 10^{-11}$ & 20.69 \\\hline
NGC 6681 & M 70& $<1.75\times 10^{-12}$ & 17.92 \\\hline
NGC 6558 & & $<1.61\times 10^{-12}$ & 2.42 \\\hline
Terzan 4 & HP 4& $<1.30\times 10^{-11}$ & 0.00 \\\hline
NGC 6715 & M 54& $<2.17\times 10^{-12}$ & 20.52 \\\hline
Arp 2 & & $<2.81\times 10^{-12}$ & 15.95 \\\hline
HP 1 & BH 229& $<5.26\times 10^{-12}$ & 22.36 \\\hline
BH 261 & AL 3& $<2.09\times 10^{-12}$ & 6.06 \\\hline
1636-283 & ESO452-SC11 & $<2.73\times 10^{-12}$ & 23.07 \\\hline
Djorg 2 & ESO456-SC38& $<2.86\times 10^{-12}$ & 3.22 \\\hline
NGC 4590 & M 68 & $<2.93\times 10^{-12}$ & 0.00 \\\hline
NGC 288 & & $<2.36\times 10^{-12}$ & 0.03 \\\hline
NGC 6293 & & $<7.33\times 10^{-12}$ & 20.31 \\\hline
NGC 5694 & & $<3.57\times 10^{-12}$ & 5.87 \\\hline
NGC 6121 & M 4 & $<1.90\times 10^{-12}$ & 6.55 \\\hline
NGC 6355 & & $<9.45\times 10^{-12}$ & 0.00 \\\hline
Terzan 10 & & $<9.44\times 10^{-12}$ & 0.00 \\\hline
NGC 6144 & & $<4.89\times 10^{-12}$ & 0.00 \\\hline
NGC 6553 & & $<1.66\times 10^{-12}$ & 1.34 \\\hline
NGC 6638 & & $<2.04\times 10^{-12}$ & 9.93 \\\hline
NGC 6284 & & $<3.67\times 10^{-12}$ & 0.00 \\\hline
NGC 6401 & & $<4.23\times 10^{-12}$ & 15.46 \\\hline
NGC 6325 & & $<7.45\times 10^{-12}$ & 1.42 \\\hline
NGC 6642 & & $<7.30\times 10^{-12}$ & 0.66 \\\hline
NGC 7099 &M 30 & $<3.44\times 10^{-12}$ & 7.64 \\\hline
Terzan 12 & & $<1.34\times 10^{-11}$ & 0.00 \\\hline
NGC 6287 & & $<6.53\times 10^{-12}$ & 5.51 \\\hline
NGC 6235 & & $<5.46\times 10^{-12}$ & 0.00 \\\hline
NGC 6864 & M 75 & $<3.06\times 10^{-12}$ & 20.48 \\\hline
Palomar 12 & & $<1.04\times 10^{-12}$ & 11.81 \\\hline
Eridanus & & $<6.93\times 10^{-13}$ & 6.00 \\\hline
NGC 5897 & & $<3.05\times 10^{-12}$ & 0.00 \\\hline
2MS-GC02 & 2MASS-GC02& $<2.97\times 10^{-12}$ & 0.59 \\\hline
Palomar 8 & & $<7.56\times 10^{-12}$ & 1.39 \\\hline
NGC 6333 & M 9& $<3.95\times 10^{-12}$ & 0.00 \\\hline
NGC 6356 & & $<4.14\times 10^{-12}$ & 0.00 \\\hline
NGC 7492 & & $<7.41\times 10^{-13}$ & 8.29 \\\hline
NGC 6171 &M 107 & $<4.21\times 10^{-12}$ & 0.00 \\\hline
NGC 6981 & M 72& $<2.81\times 10^{-12}$ & 0.00 \\\hline
Palomar 11 & & $<4.69\times 10^{-12}$ & 0.00 \\\hline
IC 1276 & Palomar 7 & $<9.51\times 10^{-12}$ & 0.00 \\\hline
NGC 5634 & & $<2.62\times 10^{-12}$ & 0.00 \\\hline
NGC 6366 & & $<1.91\times 10^{-12}$ & 0.77 \\\hline
NGC 7089 & M 2& $<2.52\times 10^{-12}$ & 0.75 \\\hline
Palomar 15 & & $<1.60\times 10^{-12}$ & 13.98 \\\hline
NGC 6535 & & $<4.04\times 10^{-12}$ & 18.22 \\\hline
Palomar 5 & & $<8.43\times 10^{-13}$ & 1.28 \\\hline
Palomar 3 & & $<2.82\times 10^{-12}$ & 2.17 \\\hline
NGC 6760 & & $<8.63\times 10^{-12}$ & 0.00 \\\hline
NGC 6749 & & $<1.01\times 10^{-11}$ & 0.00 \\\hline
NGC 6426 & & $<5.67\times 10^{-12}$ & 4.99 \\\hline
NGC 6934 & & $<2.73\times 10^{-12}$ & 0.00 \\\hline
Ko 1 & & $<2.73\times 10^{-12}$ & 0.06 \\\hline
Palomar 13 & & $<3.99\times 10^{-12}$ & 0.26 \\\hline
Palomar 14 & AvdB& $<2.68\times 10^{-12}$ & 0.01 \\\hline
NGC 7006 & & $<1.31\times 10^{-12}$ & 7.41 \\\hline
NGC 5053 & & $<2.04\times 10^{-12}$ & 17.89 \\\hline
NGC 5024 & M 53 & $<3.63\times 10^{-12}$ & 0.58 \\\hline
NGC 4147 & & $<6.24\times 10^{-13}$ & 3.16 \\\hline
Ko 2 & & $<2.96\times 10^{-12}$ & 1.22 \\\hline
NGC 5272 &M 3   & $<2.70\times 10^{-12}$ & 10.28 \\\hline
NGC 5466 & & $<2.33\times 10^{-12}$ & 6.05 \\\hline
Palomar 4 & & $<2.58\times 10^{-12}$ & 0.00 \\\hline
NGC 6779 & M 56& $<1.75\times 10^{-12}$ & 9.53 \\\hline
Palomar 2 & & $<4.40\times 10^{-12}$ & 0.00 \\\hline
NGC 6205 & M 13  & $<5.64\times 10^{-13}$ & 12.24 \\\hline
NGC 2419 & & $<2.76\times 10^{-12}$ & 0.00 \\\hline
NGC 6229 & & $<4.51\times 10^{-13}$ & 4.16 

\label{tab:low_TS}
\end{longtable}
\endgroup

\begingroup
\renewcommand\arraystretch{1.3}
\begin{longtable}{|c|c|c|c|c|}
\multicolumn{5}{c}{Table \thetable} \\
\hline
Globular Cluster & Alt.~Name & Distance~(kpc) & Luminosity~($L_{\odot}$) & $\Gamma_e$ \\
\hline
\hline
\endfirsthead

\multicolumn{5}{c}{Table \thetable\ -- continued from previous page} \\
\hline
Globular Cluster & Alt.~Name & Distance~(kpc) & Luminosity~($L_{\odot}$) & $\Gamma_e$ \\
\hline
\hline
\endhead

\multicolumn{5}{r}{continued on next page} \\
\endfoot
\hline
\noalign{\vspace{10pt}}
\captionsetup{justification=justified}
\caption{The distance, visible luminosity, and stellar encounter rate~\cite{Bahramian:2013ihw} for each of the 87 globular clusters with either $L_V > 10^5 \, L_{\odot}$ or $\Gamma_e > 0.01$. Throughout this paper, we use values of the stellar encounter rate that are normalized such that $\Gamma_e=1$ for NGC 104.}

\endlastfoot

NGC 104 & 47 Tuc & 4.46 & $5.01\times 10^{5}$ & $1.00_{-0.13}^{+0.15}$ \\
\hline
NGC 362 &  & 8.61 & $2.01\times 10^{5}$ & $0.74_{-0.12}^{+0.14}$ \\
\hline
NGC 1261 &  & 16.29 & $1.13\times 10^{5}$ & $0.015_{-0.004}^{+0.011 }$ \\
\hline
Palomar 2 &  & 27.11 & $1.32\times 10^{5}$ & $0.93_{-0.56}^{+0.84 }$ \\
\hline
NGC 1851 &  & 12.07 & $1.84\times 10^{5}$ & $1.53_{-0.19}^{+0.20 }$ \\
\hline
NGC 1904 & M79 & 12.94 & $1.19\times 10^{5}$ & $0.12_{-0.014}^{+0.019 }$ \\
\hline
NGC 2419 &  & 82.49 & $5.01\times 10^{5}$ & $0.0028^{+0.0008}_{-0.0005}$ \\
\hline
NGC 2808 &  & 9.59 & $4.88\times 10^{5}$ & $0.92_{-0.083}^{+0.067 }$ \\
\hline
NGC 4147 &  & 19.30 & $2.51\times 10^{4}$ & $0.017_{-0.006}^{+0.013}$ \\
\hline
NGC 4372 &  & 5.81 & $1.12\times 10^{5}$ & $0.0002^{+0.0004}_{-0.0001}$ \\
\hline
NGC 5024 & M53 & 17.85 & $2.61\times 10^{5}$ & $0.035_{-0.010}^{+0.012 }$ \\
\hline
NGC 5139 & Omega Centauri & 5.17 & $1.09\times 10^{6}$ & $0.090_{-0.020}^{+0.027 }$ \\
\hline
NGC 5272 & M3 & 10.18 & $3.05\times 10^{5}$ & $0.19_{-0.018}^{+0.033 }$ \\
\hline
NGC 5286 &  & 11.67 & $2.68\times 10^{5}$ & $0.46_{-0.061}^{+0.058 }$ \\
\hline
NGC 5634 &  & 25.18 & $1.02\times 10^{5}$ & $0.020_{-0.008}^{+0.014 }$ \\
\hline
NGC 5694 &  & 35.01 & $1.16\times 10^{5}$ & $0.19_{-0.034}^{+0.052 }$ \\
\hline
NGC 5824 &  & 32.17 & $2.96\times 10^{5}$ & $0.98_{-0.16}^{+0.17 }$ \\
\hline
NGC 5904 & M5 & 7.47 & $2.86\times 10^{5}$ & $0.16_{-0.030}^{+0.039 }$ \\
\hline
NGC 5927 &  & 7.67 & $1.14\times 10^{5}$ & $0.068_{-0.010}^{+0.013 }$ \\
\hline
NGC 5946 &  & 10.55 & $6.37\times 10^{4}$ & $0.13_{-0.045}^{+0.034 }$ \\
\hline
NGC 5986 &  & 10.43 & $2.03\times 10^{5}$ & $0.062_{-0.010}^{+0.016 }$ \\
\hline
NGC 6093 & M80 & 10.01 & $1.67\times 10^{5}$ & $0.53_{-0.069}^{+0.059 }$ \\
\hline
NGC 6121 & M4 & 2.22 & $6.43\times 10^{4}$ & $0.027_{-0.010}^{+0.012 }$ \\
\hline
NGC 6139 &  & 10.12 & $1.89\times 10^{5}$ & $0.31_{-0.089}^{+0.094 }$ \\
\hline
NGC 6205 & M13 & 7.14 & $2.25\times 10^{5}$ & $0.069_{-0.015}^{+0.018 }$ \\
\hline
NGC 6229 &  & 30.46 & $1.43\times 10^{5}$ & $0.048_{-0.009}^{+0.031 }$ \\
\hline
NGC 6218 & M12 & 4.83 & $7.18\times 10^{4}$ & $0.013_{-0.0040}^{+0.0054}$ \\
\hline
NGC 6254 & M10 & 4.39 & $8.39\times 10^{4}$ & $0.031_{-0.0041}^{+0.0043}$ \\
\hline
NGC 6256 &  & 10.28 & $6.19\times 10^{4}$ & $0.17_{-0.060}^{+0.119 }$ \\
\hline
NGC 6266 & M62 & 6.83 & $4.02\times 10^{5}$ & $1.67_{-0.57}^{+0.71 }$ \\
\hline
NGC 6273 & M19 & 8.80 & $3.84\times 10^{5}$ & $0.20_{-0.039}^{+0.067 }$ \\
\hline
NGC 6284 &  & 15.29 & $1.31\times 10^{5}$ & $0.67_{-0.11}^{+0.12 }$ \\
\hline
NGC 6287 & & 9.38 & $7.52\times 10^{4}$ & $0.036_{-0.0077}^{+0.0077 }$ \\
\hline
NGC 6293 & & 9.48 & $1.11\times 10^{5}$ & $0.85_{-0.24}^{+0.37 }$ \\
\hline
NGC 6304 & & 5.88 & $7.11\times 10^{4}$ & $0.12_{-0.022}^{+0.054 }$ \\
\hline
NGC 6316 & & 10.45 & $1.85\times 10^{5}$ & $0.077_{-0.015}^{+0.025 }$ \\
\hline
NGC 6341 & M92 & 8.27 & $1.64\times 10^{5}$ & $0.27_{-0.029}^{+0.030 }$ \\
\hline
NGC 6325 & & 7.83 & $5.20\times 10^{4}$ & $0.12_{-0.046}^{+0.045 }$ \\
\hline
NGC 6333 & M9 & 7.91 & $1.29\times 10^{5}$ & $0.13_{-0.042}^{+0.059 }$ \\
\hline
NGC 6342 & & 8.53 & $3.16\times 10^{4}$ & $0.045_{-0.013}^{+0.014 }$ \\
\hline
NGC 6356 & & 15.08 & $2.17\times 10^{5}$ & $0.088_{-0.014}^{+0.020 }$ \\
\hline
NGC 6355 & & 9.22 & $1.45\times 10^{5}$ & $0.10_{-0.026}^{+0.041 }$ \\
\hline
Terzan 2 & HP 3 & 7.49 & $1.92\times 10^{4}$ & $0.022_{-0.014}^{+0.029 }$ \\
\hline
NGC 6380 & Ton 1 & 10.88 & $8.55\times 10^{4}$ & $0.12_{-0.045}^{+0.068 }$ \\
\hline
NGC 6388 & & 9.92 & $4.97\times 10^{5}$ & $0.90_{-0.21}^{+0.24 }$ \\
\hline
NGC 6402 & M14 & 9.25 & $3.73\times 10^{5}$ & $0.12_{-0.030}^{+0.032 }$ \\
\hline
NGC 6401 & & 10.56 & $1.24\times 10^{5}$ & $0.044_{-0.011}^{+0.011 }$ \\
\hline
NGC 6397 & & 2.30 & $3.87\times 10^{4}$ & $0.084_{-0.018}^{+0.018 }$ \\
\hline
Palomar 6 & & 5.79 & $4.45\times 10^{4}$ & $0.016_{-0.008}^{+0.013 }$ \\
\hline
Terzan 5 & Terzan 11 & 5.98 & $7.94\times 10^{4}$ & $6.80_{-3.02}^{+1.04 }$ \\
\hline
NGC 6440 & & 8.45 & $2.70\times 10^{5}$ & $1.40_{-0.48}^{+0.63 }$ \\
\hline
NGC 6441 & & 11.60 & $6.08\times 10^{5}$ & $2.30_{-0.64}^{+0.97 }$ \\
\hline
Terzan 6 & HP 5 & 6.78 & $9.29\times 10^{4}$ & $2.47_{-1.72}^{+5.07 }$ \\
\hline
NGC 6453 & & 11.57 & $6.61\times 10^{4}$ & $0.37_{-0.09}^{+0.13 }$ \\
\hline
NGC 6517 & & 10.63 & $1.71\times 10^{5}$ & $0.34_{-0.10}^{+0.15 }$ \\
\hline
NGC 6522 & & 7.70 & $9.82\times 10^{4}$ & $0.36_{-0.10}^{+0.11 }$ \\
\hline
NGC 6528 & & 7.93 & $3.63\times 10^{4}$ & $0.28_{-0.05}^{+0.11 }$ \\
\hline
NGC 6539 & & 7.79 & $1.77\times 10^{5}$ & $0.042_{-0.015}^{+0.029 }$ \\
\hline
NGC 6544 & & 2.96 & $5.11\times 10^{4}$ & $0.11_{-0.037}^{+0.068 }$ \\
\hline
NGC 6541 & & 7.54 & $2.19\times 10^{5}$ & $0.39_{-0.063}^{+0.095 }$ \\
\hline
NGC 6553 & & 5.96 & $1.10\times 10^{5}$ & $0.069_{-0.019}^{+0.027 }$ \\
\hline
NGC 6558 & & 7.37 & $3.22\times 10^{4}$ & $0.11_{-0.019}^{+0.026 }$ \\
\hline
NGC 6569 & & 10.90 & $1.75\times 10^{5}$ & $0.054_{-0.021}^{+0.030 }$ \\
\hline
NGC 6584 & & 13.49 & $1.02\times 10^{5}$ & $0.012_{-0.0034}^{+0.0054 }$  \\
\hline
NGC 6624 & & 7.91 & $8.47\times 10^{4}$ & $1.15_{-0.18}^{+0.11 }$ \\
\hline
NGC 6626 & M28 & 5.52 & $1.57\times 10^{5}$ & $0.65_{-0.091}^{+0.084 }$ \\
\hline
NGC 6638 & & 9.41 & $6.03\times 10^{4}$ & $0.14_{-0.027}^{+0.039 }$ \\
\hline
NGC 6637 & M69 & 8.80 & $9.73\times 10^{4}$ & $0.090_{-0.018}^{+0.036 }$ \\
\hline
NGC 6642 & & 8.13 & $3.94\times 10^{4}$ & $0.10_{-0.025}^{+0.031 }$ \\
\hline
NGC 6652 & & 10.00 & $3.94\times 10^{4}$ & $0.70_{-0.19}^{+0.29 }$ \\
\hline
NGC 6656 & M22 & 3.23 & $2.15\times 10^{5}$ & $0.078_{-0.026}^{+0.032 }$ \\
\hline
NGC 6681 & M 70 & 9.01 & $6.03\times 10^{4}$ & $1.04_{-0.19}^{+0.27 }$ \\
\hline
NGC 6712 & & 6.93 & $8.55\times 10^{4}$ & $0.031_{-0.0066}^{+0.0056 }$ \\
\hline
NGC 6715 & M54 & 26.49 & $8.39\times 10^{5}$ & $2.52_{-0.27}^{+0.23 }$ \\
\hline
NGC 6717 & Palomar 9 & 7.11 & $1.57\times 10^{4}$ & $0.040_{-0.014}^{+0.022 }$ \\
\hline 
NGC 6723 & & 8.65 & $1.16\times 10^{5}$ & $0.011_{-0.0044}^{+0.0080 }$ \\
\hline
NGC 6752 & & 3.99 & $1.06\times 10^{5}$ & $0.40_{-0.13}^{+0.18 }$ \\
\hline
NGC 6760 & & 7.36 & $1.17\times 10^{5}$ & $0.057_{-0.019}^{+0.027 }$ \\
\hline
NGC 6779 & M 56 & 9.44 & $7.87\times 10^{4}$ & $0.028_{-0.009}^{+0.012 }$ \\
\hline
Palomar 10 & & 5.93 & $1.77\times 10^{4}$ & $0.059_{-0.036}^{+0.043 }$ \\
\hline
Palomar 11 & & 13.40 & $5.01\times 10^{4}$ & $0.021_{-0.007}^{+0.011 }$ \\
\hline
NGC 6864 & M 75 & 20.84 & $2.29\times 10^{5}$ & $0.31_{-0.082}^{+0.095 }$ \\
\hline
NGC 6934 & & 15.63 & $8.17\times 10^{4}$ & $0.030_{-0.008}^{+0.012 }$ \\
\hline
NGC 7006 & & 41.21 & $1.00\times 10^{5}$ & $0.0094^{+0.0049}_{-0.0033}$ \\
\hline
NGC 7078 & M 15 & 10.38 & $4.06\times 10^{5}$ & $4.51_{-0.99}^{+1.36 }$ \\
\hline
NGC 7089 & M 2 & 11.56 & $3.50\times 10^{5}$ & $0.52_{-0.071}^{+0.078 }$ \\
\hline
NGC 7099 & M 30 & 8.12 & $8.17\times 10^{4}$ & $0.32_{-0.08}^{+0.12 }$ 

\label{tab:tab7-9-hopper-linden}
\end{longtable}
\endgroup

\section{ Comparison With Previous Work}
\label{sec:app-comparison}

\begin{figure}
    \centering
    \includegraphics[width=0.5\linewidth]{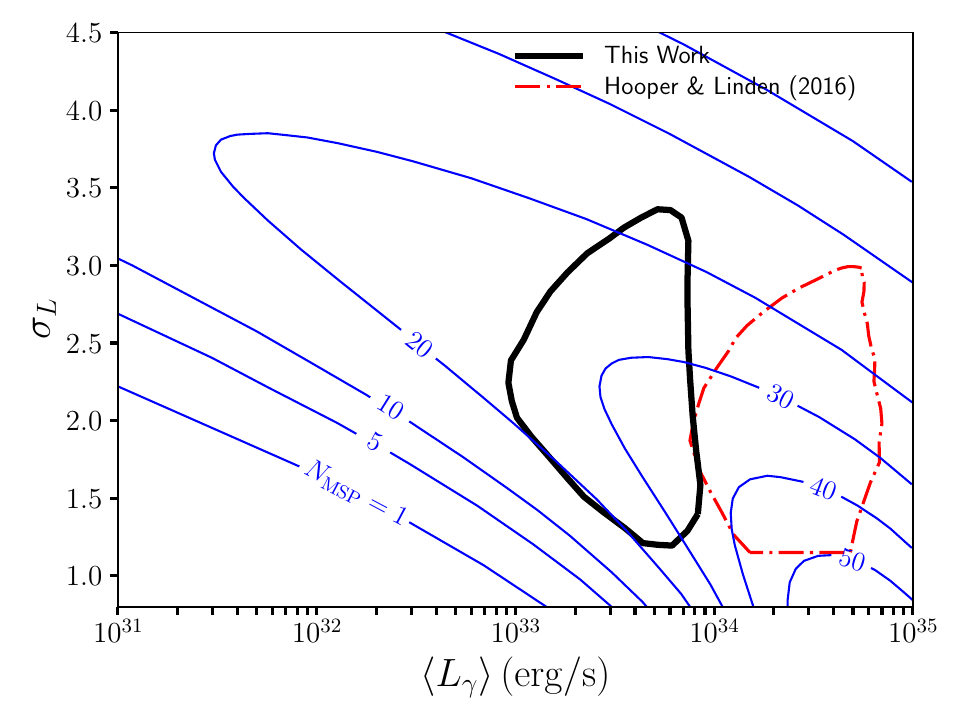}
    \caption{ As in Figure \ref{fig:3PC}, for the $2\sigma$ contours, but comparing the results from the prior 2016 work from Hooper and Linden \cite{Hooper:2016rap} with our most recent findings.}
    \label{fig:comparison_2016}
\end{figure}

In this section, we provide a direct comparison, illustrated in Figure~\ref{fig:comparison_2016}, between our results (Figure~\ref{fig:3PC}) and those from the prior 2016 work by Hooper and Linden~\cite{Hooper:2016rap}. Their analysis was based on 7 years of Fermi data, in contrast to the 15.8 years of data utilized in our work. The 2016 study favored an MSP gamma-ray luminosity function with parameters of $\langle L_{\gamma} \rangle \sim (0.7-6) \times 10^{34} \, {\rm erg/s}$ and $\sigma_L \sim 1-3$.

While our results are statistically consistent with theirs at the $\sim 2\sigma$ level, our analysis prefers smaller values for $\langle L_{\gamma} \rangle$ and larger values for $R_{\rm MSP}$ by a factor of $\sim 2-8$. Furthermore, although the best-fit values for both distributions lie near the $N_{MSP}=30$ curve, our findings suggest a narrower range of $N_{MSP} \sim 17 - 37$, compared to the previous estimate of $N_{MSP} \sim 12 - 50$. Our updated analysis refines the prior work by both encompassing the findings of Ref.~\cite{Hooper:2016rap} as far as $N_{MSP}$ is concerned, and providing tighter constraints on the number of MSPs required to explain the Galactic Center Excess.

\end{document}